\documentclass{iopart}
\usepackage{euscript,iopams,amssymb,amsfonts,graphicx,bm}
\usepackage{pgfplots,setstack}

\usepackage{float}
\usepackage{epsfig}
\usepackage{esint}

  \usepackage{paralist}
  \usepackage{epstopdf}
  \usepackage{graphics} 
 \usepackage[colorlinks=true]{hyperref}
 \hypersetup{urlcolor=blue, citecolor=red}
 \usepackage[latin1]{inputenc}
\usepackage[caption=false]{subfig}

 \usepackage{tikz-cd}

\usepackage{float}
\usepackage{epsfig}
\usepackage{esint}

\usepackage{bm,braket}

\newcommand{\pcb}[1]{\textcolor{black}{#1}}

 \newcommand{\U}{{\mathbf U}}

\renewcommand{\d}{{\rm d}}
\newcommand{\Markov}[2]{\underset{#1}{\overset{#2}{\rightleftharpoons}}}
\newcommand{\R}{{\mathbb R}}

\renewcommand{\L}{{\mathbb L}}

\renewcommand{\e}{{\rm e}}

\newcommand{\calC}{{\mathcal C}}
\newcommand{\calN}{{\mathcal N}}
\newcommand{\calP}{{\mathcal P}}
\newcommand{\calU}{{\mathcal U}}
\newcommand{\calV}{{\mathcal V}}

\newcommand{\h}{\widetilde{h}}

\newcommand{\z}{\mathbf{z}}
\newcommand{\x}{\mathbf{x}}
\newcommand{\y}{\mathbf{y}}

\newcommand{\X}{\mathbf{X}}
\renewcommand{\P}{\mathbb{P}}

\renewcommand{\u}{\mathbf u}
\newcommand{\E}{{\mathbb E}}
\newcommand{\Z}{\mathbb Z}

\eqnobysec

\begin{document}
\title[Density equations for a population of actively switching particles]{Global density equations for a population of actively switching particles}
\author{Paul C. Bressloff}
\address{Department of Mathematics, Imperial College London, London SW7 2AZ, UK.}

\date{\today}

\begin{abstract}
There are many processes in cell biology that can be modelled in terms of an actively switching particle. The continuous degrees of freedom of the particle evolve according to a hybrid stochastic differential equation (hSDE) whose drift term depends on a discrete internal or environmental state that switches according to a continuous time Markov chain. Examples include Brownian motion in a randomly switching environment, membrane voltage fluctuations in neurons, protein synthesis in gene networks, bacterial run-and-tumble motion, and motor-driven intracellular transport. \pcb{In this paper we derive global density equations for a population of non-interacting actively switching particles, either independently switching or subject to a common randomly switching environment. In the case of a random environment, we show that the global particle density evolves according to a hybrid stochastic partial differential equation (hSPDE). Averaging with respect to the Gaussian noise processes yields a hybrid partial differential equation (hPDE) for the one-particle density. We use the corresponding functional Chapman-Kolmogorov equation to derive moment equations for the one-particle density, and show how a randomly switching environment induces statistical correlations. We also discuss the effects of particle-particle interactions, which generate moment closure problems at both the hSPDE and hPDE levels. The former can be handled using dynamical density functional theory or by taking a mean field limit, but the resulting hPDE is now a nonlinear functional of the one-particle density.} We then develop the analogous constructions for independently switching particles. \pcb{We introduce a discrete set of global densities that are indexed by the single-particle internal states. We derive an SPDE for the densities in the absence of particle interactions by taking expectations with respect to the switching process, and then use a slow/fast analysis to reduce the SPDE to a scalar stochastic Fokker-Planck equation in the fast switching limit. We also indicate the difficulties in deriving a closed SPDE for the global densities when particle interactions are included.} We end the paper by deriving Martin-Siggia-Rose-Janssen-de Dominicis (MSRJD) functional path integrals for the density equations in the case of non-interacting particles, and relate this to recent field theoretic studies of Brownian gases and run-and-tumble particles (RTPs).

\end{abstract}

\maketitle

%%%%%%%%%%%%%%%%%%%%%%%%%%%%%%%%%%%%%%%%%%%%%%%%%%%%%%%

\section{Introduction}

There are a diverse range of processes in cell biology that can be modelled as an actively switching particle \cite{Bressloff17}. The continuous degrees of freedom of the particle evolve according to a hybrid stochastic differential equation (hSDE), whose drift term depends on a discrete internal or environmental state that switches according to a continuous time Markov chain. Let $(\X(t),N(t))$ denote the state of the system at time $t$ with $\X(t)\in \R^d$ and $N(t) \in \Gamma$, where $\Gamma$ is a discrete set. If $N(t)=n\in \Gamma$ then $\X(t)$ evolves according to the SDE $ d\X ={\bf A}_n(\X)dt+\sqrt{2D}d{\bf W}(t)$, where ${\bf W}(t)$ is a vector of independent Wiener processes and ${\bf A}_n$ is an $n$-dependent drift term. (For simplicity, we take the diffusivity to be independent of $n$.) Finally, $N(t)$ switches between the different discrete states according to a continuous time Markov chain whose matrix generator ${\bf Q}$ could itself depend on $\X(t)$. In the limit $D\rightarrow 0$, the dynamics reduces to a so-called piecewise deterministic Markov process \cite{Davis84}. One direct application of an hSDE is to overdamped Brownian motion in a randomly switching environment, with $\X(t)$ representing the spatial position of the particle at time $t$ and $N(t)$ the current environmental state. In particular, suppose that
 ${\bf A}_n(\X)=-\gamma^{-1} {\bm \nabla} V_n(\X)$, where $\gamma$ is a friction coefficient with $\gamma D=k_BT$ and $V_n(\X) $ is an external potential. Switching between different environmental states results in the particle being driven by different external potentials.

\begin{figure}[t!]
\centering
\includegraphics[width=14cm]{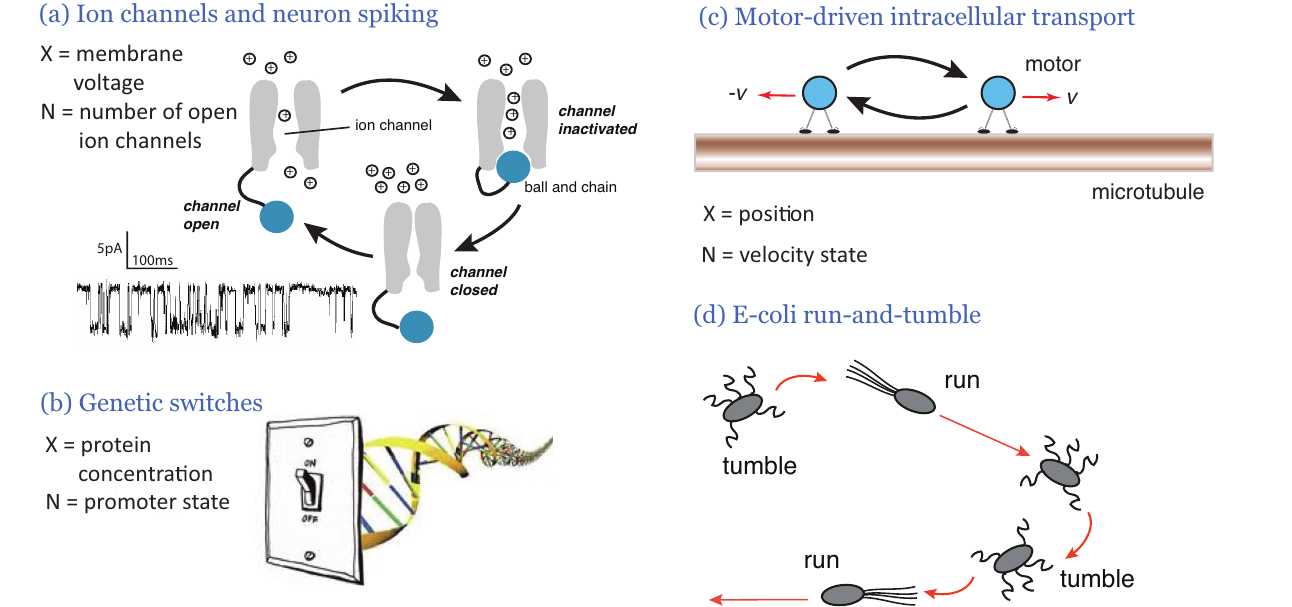} 
\caption{Examples of systems in cell biology modelled using hSDEs. (a) Ion channels. (b) Genetic switches. (c) Bidirectional motor transport. (d) Bacterial run-and-tumble.}
\label{fig1}
\end{figure}

A well-known example of an hSDE with intrinsic rather than environmental switching concerns the membrane voltage of a single neuron. In this case fluctuations in the voltage are driven by the stochastic opening and closing of ion channels \cite{Fox94,Chow96,Keener11,Goldwyn11,Buckwar11,NBK13,Bressloff14b,Newby14}, see Fig. \ref{fig1}(a). The continuous variable $X(t)\in \R$ is the membrane voltage, whereas the discrete state $N(t)$ specifies the conformational states of the ion channels (and hence the ionic membrane currents). The ion channels evolve according to a continuous-time Markov process with voltage-dependent transition rates. Applying the law of large numbers in the thermodynamic limit recovers deterministic Hodgkin-Huxley type equations. On the other hand, in the case of a finite number of channels, noise-induced spontaneous firing of a neuron can occur due to channel fluctuations. Another example of an hSDE with intrinsic switching is a gene regulatory network, where $X(t)$ is the concentration of a protein product and the discrete variable represents the activation state of the gene \cite{Kepler01,Bose04,Smiley10,Newby12,Newby15,Hufton16}, see Fig. \ref{fig1}(b). Stochastically switching between active and inactive gene states can result in translational/transcriptional bursting. Moreover, if switching persists at the phenotypic level then this provides certain advantages to cell populations growing in a changing environment, as exemplified by bacterial persistence in response to antibiotics. Yet another example is active intracellular transport on microtubular networks, where motor-cargo complexes randomly switch between different velocity states according to a special type of hSDE known as a velocity jump process \cite{Reed90,Friedman05,Newby10,Bressloff11,Bressloff13}, see Fig. \ref{fig1}(c). Velocity jump processes have also been used to model the `run-and-tumble' swimming motion of bacteria such as {\em E. coli} \cite{Berg77,Schnitzer93,Berg04}. This is characterized by periods of almost constant ballistic motion (runs), interrupted by sudden random changes in the direction of motion (tumbling), see Fig. \ref{fig1}(d). For velocity jump processes, $\X(t)$ represents the particle position and $N(t)$ specifies the velocity state.

 \pcb{Although the examples in the previous paragraph involve switching between intrinsic states, they can also be affected by environmental factors. Examples include the following: (i) an external chemical gradient regulating the tumbling rate of a run-and-tumble particle (RTP) during chemotaxis \cite{Hillen00,Erban05}; (ii) an external electric field regulating the opening and closing of a neuron's ion channels; (iii) phenotypic switching of bacterial populations in randomly switching environments \cite{Bressloff17}. Conversely, an overdamped Brownian particle can intrinsically switch between different conformational states. This may modify its effective diffusivity (possibly by temporarily binding to some some chemical substrate \cite{Das09,Persson13,Slator15}) or how it interacts with other particles (as in the case of soft colloids \cite{Bley21,Bley22}).} Mathematically speaking, at the single particle level, the analysis of an hSDE holds irrespective of whether $N(t)$ is interpreted as a discrete internal state (intrinsic switching) or an external environmental state (extrinsic switching). However, for a population of actively switching particles, the two scenarios differ significantly, even in the absence of particle-particle interactions.

In this paper, we explore the differences between intrinsic and extrinsic switching by considering a population of particles that either independently switch or are subject to a common randomly switching environment. (For simplicity, we assume throughout that the matrix generator ${\bf Q}$ is independent of $\X(t)$.) In both cases we derive `hydrodynamic' evolution equations for the global particle density by extending the classical derivation of Dean \cite{Dean96}, see also Ref. \cite{Kawasaki98}. We begin by briefly reviewing the case of a single actively switching particle, whose state $\X(t)$ evolves according to an hSDE with drift term $A_n(\X(t))$ that depends on the current discrete state $N(t)$ (section 2). We write down the differential Chapman-Kolomogorov (CK) equation for the corresponding probability density $p_n(\x,t)$, where $p_n(\x,t)d\x=\P[\X(t)\in [\x,\x+d\x,N(t)=n]$ and discuss the fast switching limit. In section 3, we consider a population of identical, non-interacting particles subject to a common randomly switching environment. \pcb{We first use Dean's construction} to derive a hybrid stochastic partial differential equation (hSPDE)
for the global density $\rho(\x,t)=\sum_{j}\delta(\X_j(t)-\x)$, where $\X_j(t)\in \R^d$ is the position of the $j$th particle at time $t$. \pcb{We show how averaging with respect to the Gaussian noise processes yields a hybrid partial differential equation (hPDE) for the one-particle density. We use the corresponding functional Chapman-Kolmogorov equation to derive moment equations for the one-particle density, and show how a randomly switching environment induces statistical correlations. We conclude section 3 by discussing how particle-particle interactions generate moment closure problems at both the hSPDE and hPDE levels. The former can be handled using dynamical density functional theory or by taking a mean field limit, but the resulting hPDE is now a nonlinear functional of the one-particle density.}
In section 4, we develop the analogous theory for a population of independently switching particles. We now introduce an indexed set of global densities defined according to $\rho_n(\x,t)=\sum_j\delta(\X_j(t)-\x)\E[\delta_{N_j(t)=n}]$, where $N_j(t)$ is the discrete state of the $j$th particle at time $t$ and expectation is taken with respect to the continuous time Markov chain. \pcb{We derive a closed non-hybrid SPDE for the densities $\rho_n(\x,t)$ in the absence of particle interactions, and show how a slow/fast analysis can be used to reduce the SPDE to a scalar stochastic Fokker-Planck equation in the fast switching limit. Deriving an SPDE for the densities $\rho_n(\x,t)$ is more complicated when particle interactions are included, since taking expectations with respect to the switching process results in a moment closure problem. We consider an application to interacting soft colloids, where a mean field ansatz is used to obtain a closed equation for the one-particle densities.} Finally, in section 5 we derive Martin-Siggia-Rose-Janssen-de Dominicis (MSRJD) functional path integrals for the density equations in the case of non-interacting particles. Throughout the paper, we emphasize the relationships between the various levels of modelling, and highlight connections with specific examples in the literature. 

\pcb{Before proceeding further, we note that the derivation of hydrodynamic equations for the one-particle density of actively switching particles
occurs in a variety of studies of active matter \cite{Tailleur08,Solon15,Zakine20,Martin21,Vrugt23}. These focus on
models of motile particles whose velocity state either randomly switches between
different discrete states (run-and-tumble motion) or undergoes rotational diffusion
(active Brownian motion). In the former case, a coarse-graining procedure is used to
approximate the single-particle dynamics by a drift-diffusion process, which is then extended to multiple interacting particles. A major application of these studies is to motility-based phase separation \cite{Cates15}.}

\section{Single actively switching particle}

Consider a particle whose states are described by a pair
of stochastic variables $(\X(t),N(t)) \in \R^d \times \{0,\cdots,K-1\}$. When the discrete state is $N(t)=n$, the particle evolves 
according to the SDE
\begin{equation}
\label{PDMP}
d\X(t)={\bf A}_{n}(\X(t))dt+\sqrt{2D}d{\bf W}(t),
 \end{equation}
where ${\bf W}$ is a vector of $d$ independent Wiener processes. The discrete stochastic variable $N(t)$ evolves according to a $K$-state
continuous-time Markov chain with a $K\times K$ matrix generator ${\bf Q}$ that is taken to be independent of $\X(t)$. It is related to the corresponding transition matrix ${\bf W}$ according to
\begin{equation}
Q_{nm} =W_{nm}-\delta_{n,m} \sum_{k=0}^{K-1}W_{km}.
\end{equation}
\pcb{We also assume that the generator is irreducible so that there exists a stationary density ${\bm \sigma}$ for which $\sum_mQ_{nm}\sigma_m=0$.}
Given the initial conditions $\X(0)=\x_0,N(0)=n_0$, we introduce the probability density $p_n(\x,t|\x_0,n_0,0) $ with
\[\P[\X(t)\in (\x,\x+d\x),\, N(t) =n|\x_0,n_0]=p_n(\x,t|\x_0,n_0,0)d\x.\]
It can be shown that $p$ evolves according to the forward differential Chapman-Kolmogorov (CK) equation \cite{Bressloff17}
\begin{equation}
\label{CKH}
\fl \frac{\partial p_n}{\partial t}=-\nabla \cdot [{\bf A}_n(\x)p_m(\x,t)]+D{\bm \nabla}^2p_n(\x,t)+\sum_{m=0}^{K-1}Q_{nm} p_{m}(\x,t).
\end{equation}
(For notational convenience we have dropped the explicit dependence on initial conditions.)
The first two terms on the right-hand side represent the probability flow associated with the SDE for a given $n$, whereas the third term 
represents jumps in the discrete state $n$. 

 In many of the applications in cell biology, one finds that the transition rates between the discrete states $m\in \{0,\ldots,K-1\}$ are much faster than the relaxation rates of the piecewise 
deterministic dynamics for \pcb{$\x\in \R^d$}. 
Thus, there is a separation of time scales between the discrete and continuous processes, so that if $t$ is the characteristic time-scale of the relaxation dynamics then 
$\epsilon t$ is the characteristic time-scale of the Markov chain for some small positive parameter $\epsilon$. Assuming that the Markov chain is ergodic, in the limit $\epsilon\rightarrow 0$ one obtains a 
deterministic dynamical system in which 
one averages the piecewise dynamics with respect to the corresponding unique stationary measure. This then raises the important problem of characterizing how the 
law of the underlying stochastic process 
approaches this deterministic limit in the case of weak noise, $0<\epsilon \ll 1$. 
Fast switching can be incorporated into the model by
introducing a small positive parameter $\epsilon$ and rescaling the transition matrix so that equation (\ref{CKH}) becomes 
\begin{equation}
\label{CKH2}
\fl \frac{\partial p_n}{\partial t}=-\nabla \cdot [{\bf A}_n(\x)p_n(\x,t)]+D{\bm \nabla}^2p_n(\x,t)+\frac{1}{\epsilon}\sum_{m=0}^{K-1}Q_{nm} p_{m}(\x,t),
\end{equation}
with ${\bf A}_n$ and ${\bf Q}$ independent of $\epsilon$ (at least to lowest order).
The fast switching limit then corresponds to the case $\varepsilon \rightarrow 0$. Let us now define the averaged vector field $\overline{A}: \R^d \to \R^d$ by
\begin{equation}
\label{Fbar}
\overline{\bf A}(\x)=\sum_{m =0}^{K-1}\sigma_m {\bf A}_m(\x).
\end{equation}
Intuitively speaking, one would expect the hSDE (\ref{PDMP}) to reduce to the classical SDE
\begin{equation}
\label{mft}
\d\X(t) =    \overline{\bf A}(\X(t))dt+\sqrt{2D}d{\bf W}(t)\end{equation}
 in the fast switching limit $\varepsilon \rightarrow 0$. This can be made precise in terms of a law of large numbers for stochastic 
hybrid systems \cite{Kifer09,fagg09,fagg10,Bressloff17a}. \pcb{The Fokker-Planck equation corresponding to the SDE (\ref{mft}) is obtained by taking $\rho_n(\x,t)=\sigma_n p(\x,t)+O(\epsilon)$ in equation (\ref{CKH2}), summing with respect to $n$, and noting that $\sum_nQ_{nm}=0$ and $\sum_n\sigma_n=1$.}

In the case of the hSDE (\ref{PDMP}) for the pair $(\X(t),N(t))$, the stochastic dynamics is the same whether $N(t)$ is interpreted as a discrete internal state or an external environmental state. However, for a population of non-interacting actively switching particles, the two scenarios differ significantly. First, suppose that there are $\calN$ particles with continuous states $\X_j(t)$, $j=1,\ldots,\calN$, and subject to independent white noise processes ${\bf W}_j(t)$. If each particle has its own internal state $N_j(t)$, then the population version of equation (\ref{PDMP}) is
\begin{equation}
\label{multiPDMPi}
d\X_j(t)={\bf A}_{n_j}(\X_j(t))dt+\sqrt{2D}d{\bf W}_j(t),\quad N_j(t)=n_j.
 \end{equation}
 We can treat the full system as the product of $\calN$ independent hSDEs $(\X_j(t),N_j(t))$ (assuming that the initial conditions are also independent).
 On the other hand, if each particle is subject to the same discrete environmental state $N(t)$ then
\begin{equation}
\label{multiPDMPd}
d\X_j(t)={\bf A}_{n}(\X_j(t))dt+\sqrt{2D}d{\bf W}_j(t),\quad N(t)=n.
 \end{equation}
Since $N(t)$ is a global variable that is experienced by all members of the population, it follows that the particles are statistically correlated, even in the absence of particle-particle interactions. One way to interpret equation (\ref{multiPDMPd}) is as a single hSDE for the high-dimensional system $({\bf Z}(t),N(t))$ with ${\bf Z}(t)=(\X_1(t),\ldots,\X_{\calN}(t))$. The corresponding CK equation for the probability density $\calP_n({\bf Z},t)$ takes the form
 \begin{eqnarray}
\label{mbDL}
\fl  \frac{\partial \calP_{n}}{\partial t}&=- \sum_{j=1}^{\calN}\nabla_j\cdot [{\bf A}_n(\x_j)\calP_{n}({\bf z},t)]+D\sum_{j=1}^{\calN}\nabla_j^2\calP_{n}(\z,t) +\sum_{m=0}^{K-1}Q_{nm}\calP_{m}(\z,t),
 \end{eqnarray}
 where $\nabla_j=\nabla_{\x_j}$ and $N(t)=n$.
 The presence of the final term on the right-hand side means that we cannot factorize the solution according to 
 $\calP_n(\z,t)=\prod_{j=1}^{\calN}P_n(\x_j,t)$, where $P_n(\x,t)$ satisfies the CK equation (\ref{CKH}). This reflects the existence of statistical correlations. In the following sections we explore these differences in terms of hydrodynamical equations for the global particle density.

\section{Population of particles with environmental switching}

\subsection{Global density \pcb{(non-interacting particles)}} Consider a population of non-interacting particles in the presence of a randomly switching environment. \pcb{A typical example would be a population of overdamped Brownian particles subject to a common randomly switching external potential.} A more compact description of the population dynamics of equation (\ref{multiPDMPd}) can be obtained by considering a ``hydrodynamic'' formulation that involves the
 global density
\begin{equation}
\rho(\x,t)=\sum_{j=1}^{\calN}\rho_j(\x,t),\quad \rho_j(\x,t)=\delta(\X_j(t)-\x).
\end{equation}
In particular, following along identical lines to Dean \cite{Dean96}, we can derive an Ito SPDE for $\rho(\x,t)$ that depends on the environmental state via the drift vector ${\bf A}_n$. For completeness, we sketch the basic steps. Suppose that at time $t$ the environmental state is $N(t)=n$. 
Consider an arbitrary smooth function $f: \R^d\rightarrow \R$.
 Using Ito's lemma to Taylor expand $f(\X_i(t+dt))$ about $\X_i(t)$ and setting $
f(\X_i(t))=\int_{\R^d} \rho_i(\x,t)f(\x)d\x$, we find that
\begin{eqnarray}
\fl \frac{df(\X_i)}{dt}&=\int_{\R^d}d\x\, f(\x)\frac{\partial \rho_i(\x,t)}{\partial t}\\
\fl &=\int_{\R^d}d\x\, \rho_i(\x,t)\bigg [\sqrt{2D}{\bm \nabla} f(\x)\cdot {\bm \xi}_i(t)+D{\bm \nabla}^2 f(\x)+{\bm \nabla} f(\x)\cdot {\bf A}_n(\x)\bigg ],\  N(t)=n. \nonumber 
\end{eqnarray}
We have formally set $d{\bf W}_i(t)={\bm \xi}_i(t)dt$ where $ {\bm \xi}_i $ is a $d$-dimensional white noise term such that
\begin{equation}
\langle {\bm \xi}_i(t)\rangle =0,\quad \langle {\xi}_i^{\sigma}(t){\xi}_j^{\sigma'}(t')\rangle =\delta(t-t')\delta_{i,j}\delta_{\sigma,\sigma'}.
\end{equation}
Integrating by parts the various terms on the second line and using the fact that $f$ is arbitrary yields a PDE for $\rho_i$:
\begin{eqnarray}
\fl \frac{\partial \rho_i(\x,t)}{\partial t} 
&=-\sqrt{2D}{\bm \nabla} \cdot [ \rho_i(\x,t) {\bm \xi}_i(t)]+D{\bm \nabla}^2  \rho_i(\x,t)-{\bm \nabla} \cdot [ \rho_i(\x,t){\bf A}_n(\x)]
\end{eqnarray}
for $N(t)=n$.
Summing over the particle index $i$ and using the definition of the global density then gives
\begin{eqnarray}
\fl \frac{\partial \rho(\x,t)}{\partial t} 
 &=-\sqrt{2D}\sum_{i=1}^{\calN}{\bm \nabla} \cdot \bigg [ \rho_i(\x,t) {\bm \xi}_i(t)\bigg ]+D{\bm \nabla}^2  \rho(\x,t)-{\bm \nabla} \cdot [ \rho(\x,t){\bf A}_n(\x)].
\label{rho1}
\end{eqnarray}
As it stands, equation (\ref{rho1}) is not a closed equation for $\rho$ due to the noise terms. Following Ref. \cite{Dean96},  we introduce the space-dependent Gaussian noise term
\begin{equation}
{ \xi}(\x,t)=-\sum_{i=1}^{\calN}{\bm \nabla} \cdot \bigg [ \rho_i(\x,t) {\bm \xi}_i(t)\bigg ]
\end{equation}
with zero mean and the correlation function
\begin{equation}
\langle 
\xi (\x,t)\xi (\y,t')\rangle = \delta(t-t')\sum_{i=1}^{\calN} {\bm \nabla}_{\x}\cdot {\bm \nabla}_{\y}\bigg (\rho_i(\x,t) \rho_i(\y,t) \bigg ).
\end{equation}
Since $\rho_i(\x,t) \rho_i(\y,t) =\delta(\x-\y)\rho_i(\y,t)$, it follows that
\begin{equation}
\langle 
\xi (\x,t)\xi (\y,t')\rangle = \delta(t-t') {\bm \nabla}_{\x}\cdot {\bm \nabla}_{\y}\bigg (\delta(\x-\y)\rho(\x,t) \bigg ).
\end{equation}
Finally, we introduce the global density-dependent noise field
\begin{equation}
\widehat{\xi }(\x,t)={\bm \nabla} \cdot\bigg ({\bf \eta}(\x,t)\sqrt{\rho}(\x,t)\bigg ),
\end{equation}
where ${\bf \eta}(\x,t)$ is a global white noise field whose components satisfy 
\begin{equation}
\langle  \eta^{\sigma}(\x,t)\eta^{\sigma'}(\y,t')\rangle =\delta(t-t')\delta(\x-\y)\delta_{\sigma,\sigma'}.
\end{equation}
It can be checked that the Gaussian noises ${ \xi}$ and $\widehat{\xi}$ have the same correlation functions and are thus statistically identical. We thus obtain a closed hSPDE for the global density that couples to the environmental state $N(t)=n$:
\begin{eqnarray}
\fl \frac{\partial \rho(\x,t)}{\partial t} 
&=\sqrt{2D}{\bm \nabla} \cdot \bigg [ \sqrt{\rho(\x,t)} {\bm \eta}(\x,t)\bigg ]+D{\bm \nabla}^2  \rho(\x,t) 
-{\bm \nabla} \cdot  [\rho(\x,t){\bf A}_n(\x)] . 
\label{rhoc}\end{eqnarray}
Note that the $n$-dependence of the drift vector ${\bf A}_n$ introduces another level of stochasticity due to the randomly switching environment. As we show below, this introduces statistical correlations between the particles. On the other hand, if the drift term is independent of the environmental state, then statistical correlations will only occur if there are particle-particle interactions \cite{Dean96}, see also section 3.3.

\subsection{Statistical correlations and moment equations for the one-body density}

In order to investigate statistical correlations induced by the random environment, 
we average the hSPDE (\ref{rhoc}) with respect to the independent Gaussian noise terms to obtain a closed hPDE for the one-body density
\begin{equation}
u(\x,t)=\langle \rho(\x,t)\rangle .
\end{equation}
(If pairwise particle interactions were included then $\langle \rho(\x,t)\rangle$ would couple to the second order moment $\langle \rho(\x,t)\rho(\y,t)\rangle$ etc. Hence, moment closure would no longer hold \cite{Dean96}, see below.)
Between jumps in the environmental state, the density $u(\x,t)$ evolves according to the drift-diffusion equation
\begin{eqnarray}
\label{pc}
&\frac{\partial u(\x,t)}{\partial t} =D{\bm \nabla}^2  u(\x,t)-{\bm \nabla} \cdot \bigg ( u(\x,t){\bf A}_n(\x)\bigg ) ,
\end{eqnarray}
This type of stochastic hybrid model can be analyzed along similar lines to reaction-diffusion equations with randomly switching boundaries \cite{Bressloff15a,Bressloff15b,Bressloff16}. We proceed by spatially discretizing equation (\ref{pc}) in terms of a $d$-dimensional regular lattice with nodes ${\bm \ell} \in \Z^d$ and lattice spacing $h$. Let $\Gamma_{{\bm \ell}}$ denote the set of nearest neighbors of ${\bm \ell}$:
\begin{equation}
\Gamma_{{\bm \ell}}=\{{\bm \ell} \pm h{\bf e}_{\sigma}, \sigma=1,\ldots d\},
\end{equation}
where ${\bf e}_{\sigma}$ is the unit vector along the $\sigma$-axis.
Setting $u_{{\bm \ell}}(t)=u({\bm \ell} h,t)$ \pcb{and $A_{n,{\bm \ell}}^{\sigma}=A_n^{\sigma}({{\bm \ell}} h)$}, ${\bm \ell}\in {\mathbb Z}^d$, we obtain the piecewise deterministic ODE
\begin{equation}
\label{sh}
\frac{du_{{\bm \ell}}}{dt}=D[\Delta(\u)]_{{\bm \ell}} -\frac{1}{h}\sum_{\sigma=1}^d\pcb{[u_{{\bm \ell}+{\bf e}_{\sigma}}A^{\sigma}_{n,{\bm \ell}+{\bf e}_{\sigma} }-u_{{\bm \ell}}A_{n,{\bm \ell}}^{\sigma}]}
\end{equation}
for ${\bm \ell} \in {\mathbb Z}^d$, where $\Delta$ is the discrete Laplacian 
\begin{equation}
\label{dL1}
[\Delta(\u)]_{{\bm \ell}}=\frac{1}{h^2} \sum_{{\bm \ell}'\in \Gamma_{{\bm \ell}}}[u_{{\bm \ell}'}-u_{{\bm \ell}}] .
\end{equation}
Introducing the infinite-dimensional vector $\U=(u_{{\bm \ell}},\, {\bm \ell} \in \Z^d)$ and the corresponding probability density
\begin{equation}
{\mathcal P}_n(\u,t)d\u= \P\{\U(t)\in (\u,\u+d\u), N(t)=n\},
\end{equation}
 we can now write down a CK equation for the spatially discretized system: 
\begin{eqnarray}
\fl  \frac{\partial \calP_n}{\partial t}&=-\sum_{{\bm \ell}\in \Z^d}\frac{\partial}{\partial u_{{\bm \ell}}}\left [\left (D[\Delta(\u)]_{{\bm \ell}} -\frac{1}{h}\sum_{\sigma=1}^d\pcb{[u_{{\bm \ell}+{\bf e}_{\sigma}}A^{\sigma}_{n,{\bm \ell}+{\bf e}_{\sigma} }-u_{{\bm \ell}}A_{n,{\bm \ell}}^{\sigma}]}\right )\calP_n(\u,t)\right ] \nonumber \\
\fl &\quad +\sum_{m }Q_{nm}\calP_m(\u,t).
\label{swCK00}
\end{eqnarray}
Finally, we take the continuum limit $h\rightarrow 0$ of equation (\ref{swCK00}). This  
yields a functional CK equation for the many-body probability functional $\calP_n[u,t]$ with
\begin{equation}
\calP_n[u,t]\prod_{\x}du(\x)=\lim_{h\rightarrow 0}\calP_n[{\bf u},t]d\u.
\end{equation}
That is,
\begin{eqnarray}
\fl  \frac{\partial \calP_n[u,t]}{\partial t}&=-\int_{\R^d} d\x\, \frac{\delta}{\delta u(\x,t)}\bigg [\left (D{\bm \nabla}^2u(\x,t)-{\bm \nabla}\cdot [u(\x,t){\bf A}_n(\x)]\right ) \calP_n[u,t]\bigg ]\nonumber \\
\fl  &\quad +\sum_{m }Q_{nm}\calP_m[u,t].
 \label{swCK0}
\end{eqnarray}

Equation (\ref{swCK0}) can now be used to derive various moment equations. For example, the first moment is defined as
\begin{eqnarray}
\label{mfirst}
{\mathcal U}_n(\x_1,t)\equiv \E[u(\x_1,t) 1_{N(t)=n}]=\int D[u]\, u(\x_1)\calP_n[u,t].
\end{eqnarray}
We use $\E[\cdot]$ to denote expectation with respect to the switching process, in order to contrast it with $\langle \cdot \rangle$, which denotes averaging with respect to the Gaussian noise, that is, $u(\x,t)=\langle \rho(\x,t)\rangle$ etc. Multiplying both sides of equation (\ref{swCK0}) by $u(\x_1)$ and functionally integrating with respect to $u$ yields the first moment equation 
\begin{eqnarray}
\label{mom1}
\fl   \frac{\partial {\mathcal U}_n}{\partial t}&=D{\bm \nabla}^2{\mathcal U}_n(\x_1,t)-{\bm \nabla}\cdot  {\mathcal U}_n(\x_1,t){\bf A}_n(\x_1) + \sum_{m }Q_{nm}{\mathcal U}_{m}(\x_1,t).
\end{eqnarray}
Note that the first moment equation (\ref{mom1}) is identical to the CK equation (\ref{CKH}) for the single particle hSDE (\ref{PDMP}) under the mapping $P_n(\x,t)\rightarrow {\mathcal U}_n(\x,t)=\calN P_n(\x,t)$.
On the other hand, the second-order moments
\begin{equation}
C_n(\x_1,\x_2,t)=\E[u(\x_1,t)u(\x_2,t)1_{N(t)=n}].
\end{equation}
evolves according to the moment equation
  \begin{eqnarray}
   \label{C22}
  \fl \frac{\partial C_n}{\partial t} &=D{\bm \nabla}_{1}^2 C_n(\x_1,\x_2,t)+D{\bm \nabla}_{2}^2C_n(\x_1,\x_2,t)\\
 \fl  & -{\bm \nabla}_{1}\cdot C_n(\x_1,\x_2,t){\bf A}_n(\x_1) -{\bm \nabla}_{2}\cdot C_n(\x_2,\x_2,t){\bf A}_n(\x_2) 
  +\sum_{m }Q_{nm}C_m(\x_1,\x_2,t).\nonumber 
  \end{eqnarray}
The latter can be derived from equation (\ref{swCK0}) after multiplying both sides by the product $u(\x,1)u(\x_2)$ and functionally integrating by parts. Clearly $C_n(\x_1,\x_2,t)\neq \calU_n(\x_1,t)\calU_n(\x_2,t)$, which means that the two-point correlation function is non-zero. A similar comment holds for higher-order moments. Interestingly, the second moment equation (\ref{C22}) takes the form of a CK equation for an effective single particle hSDE with $2d$ continuous coordinates $(\X(t),{\bf Y}(t))$: if $N(t)=n$ then
\numparts
\begin{eqnarray}
\label{2mom}
d\X(t)&={\bf A}_n(\X(t))dt+\sqrt{2D}d{\bf W}_1(t),\\
d{\bf Y}(t)&={\bf A}_n({\bf Y}(t))dt+\sqrt{2D}d{\bf W}_2(t),
 \end{eqnarray}
 \endnumparts
where $({\bf W}_1,{\bf W}_2)^{\top}$ is a vector of $2d$ independent Wiener processes.

\begin{figure}[t!]
 \centering
\includegraphics[width=13cm]{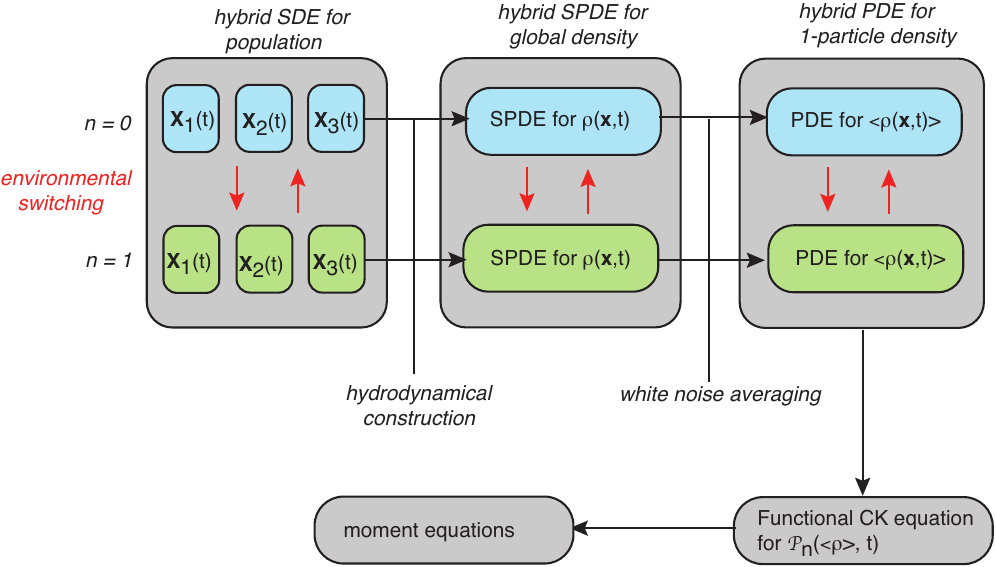}
\caption{\pcb{Hierarchy of hybrid models for the density function of a population of non-interacting particles in a randomly switching environment. For the sake of illustration, we assume that there are two environmental states labelled $n=0,1$.}}\label{fig2}
\end{figure}

Finally, note that we could extend the above analysis to the full global density $\rho(\x,t)$. For example, the corresponding functional CK equation would now take the form
\begin{eqnarray}
\fl  \frac{\partial \calP_n[\rho,t]}{\partial t}&=-\int_{\R^d} d\x\, \frac{\delta}{\delta \rho(\x,t)}\bigg [\left (D{\bm \nabla}^2\rho(\x,t)-{\bm \nabla}\cdot [\rho(\x,t){\bf A}_n(\x)]\right ) \calP_n[\rho,t]\bigg ]\nonumber \\
\fl  &\quad +D\int_{\R^d} d\x {\bm \nabla}^2 \frac{\delta^2}{\delta \rho(\x,t)^2}\bigg [ \rho(\x,t)\calP_n[\rho,t]\bigg ]+\sum_{m }Q_{nm}\calP_m[\rho,t].
 \label{swCK0A}
\end{eqnarray}
The term involving the second-order functional derivative does not contribute to the first-moment equation and generates a spatially local term in the second-order moment equation. That is, setting
\begin{equation}
\calC_n(\x,\y,t)=\E[\langle \rho(\x,t)\rho(\y,t)1_{N(t)=n}\rangle ],
\end{equation}
which averages with respect to the Gaussian noise and the random switching,
we find that
\begin{eqnarray}
  \label{calC22}
  \fl \frac{\partial \calC_n}{\partial t} &=D{\bm \nabla}_{\x}^2 \calC_n(\x,\y,t)+D{\bm \nabla}_{\y}^2\calC_n(\x,\y,t)+2D{\bm \nabla}^2 [u(\x,t)\delta(\x-\y)] \nonumber \\
 \fl  &\quad -{\bm \nabla}_{\x}\cdot \calC_n(\x,\y,t){\bf A}_n(\x) -{\bm \nabla}_{\y}\cdot \calC_n(\x,\y,t){\bf A}_n(\y) 
  +\sum_{m }Q_{nm}\calC_m(\x,\y,t).
  \end{eqnarray}
   In Fig. \ref{fig2} we summarize the various versions of the population model in the case of environmental switching.

\subsection{Interacting Brownian particles and mean field theory} \pcb{So far we have ignored the effects of particle interactions. In the case of overdamped Brownian particles, such interactions are typically taken to be pair-wise so that equation (\ref{multiPDMPd}) becomes}
\begin{equation}
\fl d\X_j(t)={\bf A}_{n}(\X_j(t))dt+\sum_{k=1}^{\calN} {\bf F} (\X_j(t)-\X_k(t))+\sqrt{2D}d{\bf W}_j(t),\quad N(t)=n.
\label{mooenv}
 \end{equation}
 \pcb{If the forces are conservative, then ${\bf A}_n(\x)=-\beta D{\bm \nabla}V_n(\x)$ and ${\bf F}(\x)=-\beta D{\bm \nabla}K(\x)$, $\beta=1/k_BT$, where $V_n$ is an environment-dependent potential and $K$ is an interaction potential. The latter could also depend on the environmental state $N(t)$.} The derivation of equation (\ref{rhoc}) can be extended to include particle interactions along analogous lines to Ref. \cite{Dean96}. The global density now evolves according to the hSPDE  
 \begin{eqnarray}
\fl \frac{\partial \rho(\x,t)}{\partial t} 
&=\sqrt{2D}{\bm \nabla} \cdot \bigg [ \sqrt{\rho(\x,t)} {\bm \eta}(\x,t)\bigg ]+D{\bm \nabla}^2  \rho(\x,t)\nonumber \\
\fl &\quad  -{\bm \nabla} \cdot  \rho(\x,t)\bigg ( {\bf A}_n(\x)+\int_{\R^d}\rho(\y,t) {\bf F}(\x-\y)d\y\bigg ) ,\quad N(t)=n. 
\label{rhoc2}\end{eqnarray}
When particle interactions are included, averaging equation (\ref{rhoc2}) with respect to the Gaussian noise no longer generates a closed equation for $u(\x,t)=\langle \rho(\x,t)\rangle$:
\begin{eqnarray}
\frac{\partial u(\x,t)}{\partial t} &=D{\bm \nabla}^2    u(\x,t) -{\bm \nabla} \cdot  [u(\x,t) {\bf A}_n(\x)]\nonumber \\
&\quad -{\bm \nabla} \cdot  \int_{\R^d} \langle\rho(\x,t) \rho(\y,t)\rangle {\bf F}(\x-\y)d\y,\quad N(t)=n  . 
\label{rhocU}
\end{eqnarray}
(We are using the fact that the Gaussian noise is independent of $N(t)$.)
That is, $u(\x,t)$ couples to the two-point correlation function
\begin{equation}
u^{(2)}(\x,\y,t)=\langle\rho(\x,t) \rho(\y,t)\rangle,
\end{equation}
which, in turn, depends on the three-point correlation function etc.

One way to achieve moment closure for the one-body density is to use dynamical density functional theory (DDFT) \cite{Marconi99,Evans04,Archer04,Witt21}. 
A crucial assumption of DDFT is that the relaxation of the system is sufficiently slow such that the pair correlation can be equated with that of a corresponding equilibrium system at each point in time \cite{Witt21}. This allows one to approximate equation (\ref{rhocU}) by the closed hPDE
\begin{equation}
\frac{\partial u(\x,t)}{\partial t}=-{\bm \nabla}\cdot {\bf J}_n(\x,t), \quad N(t)=n
\label{DDFT0}
\end{equation}
where
\begin{eqnarray}
{\bf J}_n(\x,t)=-D\bigg \{{\bm \nabla} u(\x,t)+\beta u(\x,t){\bm \nabla} [V_n(\x)+\mu^{\rm ex}(\x,t)]\bigg \},
\end{eqnarray}
Here 
\begin{equation}
\mu^{\rm ex}(\x,t)=\frac{\delta F^{\rm ex}[u(\x,t)]}{\delta u(\x,t)}
\end{equation}
and $F^{\rm ex}[u]$ is the equilibrium excess free energy functional with the equilibrium density profiles replaced by non-equilibrium ones. One of the features of DDFT is that $F^{\rm ex}[u]$ is independent of the actual external potential, and is thus independent of the environmental state.

\pcb{In order to apply DDFT, it is necessary to take account of the fact that there is another time-scale, namely, the rate of environmental switching. 
How this affects the validity of the adiabatic approximation remains to be determined, and is the subject of future work. Intuitively speaking, in the fast switching limit one could first average with respect to the switching process along analogous lines to an hSDE, see equation (\ref{mft}), and then apply DDFT. That is, $A_n(\x)$ in equation (\ref{rhoc2}) would be replaced by $\overline{A}(\x)=\sum_m \sigma_m A_m(\x)$. On the other hand, in the slow switching limit one could apply DDFT for fixed $n$ and then consider switching between the $n$-dependent 1-particle density equations. However, since the hPDE (\ref{DDFT0}) for fixed $n$ takes the form of a nonlinear Fokker-Planck equation, it is no longer possible to obtain closed moment equations using the corresponding functional CK equation for $ \calP_n[u,t]$. Even assuming that both limiting cases are well-posed, it is unclear what would happen at intermediate switching rates. }

\pcb{An alternative way of deriving a closed hPDE for an effective one-particle density is to use mean field theory. In the case of non-switching, weakly-interacting Brownian particles, there is an extensive mathematical literature on the mean field limit (or propagation of chaos), see the review \cite{Jabin17} and Refs. \cite{Oelsch84,Carrillo20}. More specifically, suppose that both the external and interaction potentials are independent of the environmental state, and take $K= K_0/\calN$ where $K_0$ is a smooth function. Equation (\ref{mooenv}) then takes the form
\begin{eqnarray}
\fl d\X_j(t)=-D\beta  \pcb{{\bm \nabla}} V(\X_j(t))dt-\frac{D\beta}{\calN}\sum_{k=1}^{\calN} {\bm \nabla}K_0 (|\X_j(t)-\X_k(t)|)+\sqrt{2D}d{\bf W}_j(t).
\label{moo2}
 \end{eqnarray}
 Introducing the normalized global density (or empirical measure)
 \begin{equation}
  \rho_{\calN}(\x,t)=\frac{1}{\calN}\sum_{j=1}^{\calN}\delta(\x-\X_j(t)),
 \end{equation}
 the Dean-Kawasaki equation becomes
  \begin{eqnarray}
\fl  \frac{\partial \rho_{\calN}(\x,t)}{\partial t} 
&=\sqrt{\frac{2D}{\calN}}{\bm \nabla} \cdot \bigg [ \sqrt{\rho_{\calN}(\x,t)} {\bm \eta}(\x,t)\bigg ]+D{\bm \nabla}^2  \rho_{\calN}(\x,t)  \\
\fl & \quad  +D\beta {\bm \nabla} \cdot   \rho_{\calN}(\x,t)\bigg ({\bm \nabla}V(\x)+\int_{\R^d}  \rho_{\calN}(\y,t) {\bm \nabla}K_0(|\x-\y|)\d\y \bigg ) .\nonumber
\label{DKN}
\end{eqnarray}
Furthermore, suppose that the joint probability density at $t=0$ takes the product form
\begin{equation}
p(\x_1,\ldots,\x_{\calN},0)=\prod_{j=1}^{\calN}\rho_0(\x_j).
\end{equation}
It can then be proven that, as $\calN\rightarrow \infty$, $\E[\rho_{\calN}]$ converges in distribution to the solution $\rho$ of the so-called McKean-Vlasov equation \cite{McKean66}
\begin{eqnarray}
\fl \frac{\partial \rho(\x,t)}{\partial t} 
&=D{\bm \nabla}^2  \rho(\x,t)  +D\beta {\bm \nabla} \cdot  \rho(\x,t)\bigg ( {\bm \nabla}V(\x) +\int_{\R^d}  \rho(\y,t) {\bm \nabla}K_0(|\x-\y|)d\y\bigg ),
\label{MV}\end{eqnarray}
with $\rho(\x,0)=\rho_0(\x)$. Recently, this classical result has been extended to the case of weakly-interacting Brownian particles in a common randomly switching environment \cite{Nguyen20}.
}

 \section{Population of particles with intrinsic switching}

\subsection{Global density \pcb{(non-interacting particles)}}

The derivation of an evolution equation for the global density differs significantly for a population of particles that independently switch between a set of internal states \pcb{along the lines of equation (\ref{multiPDMPi}). Examples include regulatory gene networks, run-and-tumble particles, soft colloids, and molecular motors.}  First, we modify the definition of the global density by setting
\begin{equation}
 {\rho}_n(\x,t)=\sum_{j=1}^{\calN}\rho_j(\x,t)\E[\delta_{N_j(t),n}],\quad \rho_j(\x,t)=\delta(\X_j(t)-\x).
 \label{dglob}
\end{equation}
In order to derive an equation for $\rho_n$, we introduce an arbitrary set of smooth functions $f_n(x)$ such that
\begin{equation}
f_{N_j(t)}(\X_j(t))=\sum_{n=0}^{K-1} \int_{\R^d}\rho_j(\x,t)\delta_{N_j(t),n}f_n(\x)d\x.
\label{fN}
\end{equation}
We then note that
\begin{eqnarray}
\fl & f_{N_i(t+\Delta t)}(\X_i(t+\Delta t))=\sum_{n=0}^{K-1} f_{n}(\X_i(t+\Delta t)) \delta_{n,N_i(t+\Delta t)}\nonumber \\
\fl &=\sum_{n =0}^{K-1}\bigg [f_{n}(\X_i(t))+{\bm \nabla}f_n(\X_i(t))\cdot \Delta \X_i(t)+\ldots \bigg] \delta_{n,N_i(t+\Delta t)}.
\end{eqnarray}
Applying Ito's lemma we have
\begin{eqnarray}
\fl  & f_{N_i(t+\Delta t)}(\X_i(t+\Delta t))\nonumber \\
\fl &\approx \sum_{n=0}^{K-1}\delta_{n,N_i(t+\Delta t)}\int_{\R^d}d\x\, \rho_i(\x,t)\bigg \{f_n(\x)+\Delta t\bigg [\sqrt{2D}{\bm \nabla} f_n(\x)\cdot {\bm \xi}_i(t)+D{\bm \nabla}^2 f_n(\x)\nonumber 
\\
\fl & \hspace{4cm}+{\bm \nabla} f_n(\x)\cdot {\bf A}_n(\x)\bigg ]+O(\Delta t^2)\bigg \} \nonumber \\
\fl &=f_{N_i(t)}(X_i(t)) \nonumber \\
\fl &+\Delta t\sum_{n=0}^{K-1}\delta_{n,N_i(t)} \int_{\R^d}d\x\, \rho_i(\x,t)\bigg [\sqrt{2D}{\bm \nabla} f_n(\x)\cdot {\bm \xi}_i(t)+D{\bm \nabla}^2 f_n(\x)+{\bm \nabla} f_n(\x)\cdot {\bf A}_n(\x)\bigg ] \nonumber \\
\fl &\hspace{2cm}+\sum_{n=0}^{K-1}[\delta_{n,N_i(t+\Delta t)}-\delta_{n,N_i(t)}]\int_{\R^d}d\x\, \rho_i(\x,t)\bigg \{f_n(\x)+O(\Delta t)\bigg \}.
\end{eqnarray}
Rearranging this equation, dividing through by $\Delta t$ and taking the limit $\Delta t \rightarrow 0$ gives
\begin{eqnarray}
\fl &\frac{d f_{N_i(t)}(\X_i(t))}{d t} 
=\delta_{n,N_i(t)}\int_{\R^d}d\x\, \rho_i(\x,t)\bigg [\sqrt{2D}{\bm \nabla} f_n(\x)\cdot {\bm \xi}_i(t)+D{\bm \nabla}^2 f_n(\x)\nonumber \\
\fl &\qquad +{\bm \nabla} f_n(\x)\cdot {\bf A}_n(\x)\bigg ]+ \lim_{\Delta t \rightarrow 0} \sum_{n=0}^{K-1}\frac{\delta_{n,N_i(t+\Delta t)}-\delta_{n,N_i(t)}}{\Delta t}\int_{\R^d}d\x\, \rho_i(\x,t) f_n(\x).
\end{eqnarray}

Substituting for $ f_{N_i(t)}(\X_i(t))$ on the left-hand side using equation (\ref{fN}), we then take expectations with respect to the Markov chain. Setting $ {\rho}_i(\x,n,t)=\rho_i(\x,t)\E[\delta_{n,N_i(t)}]$ and using the fact that\begin{equation}
\E\bigg [\lim_{\Delta t \rightarrow 0}\frac{\delta_{n,N_i(t+\Delta t)}-\delta_{n,N_i(t)}}{\Delta t} \bigg ]=\sum_{m=0}^{K-1}Q_{nm}\E[\delta_{m,N_i(t)}],
\label{mum}
\end{equation}
where ${\bf Q}$ is the matrix generator\footnote{The numerator in equation (\ref{mum}) is equal to 1 if $N_i(t)\neq n$ and $N_{i}(t+\Delta t) =n$, that is, there is a transition $m\rightarrow n$ for some $m\neq n$ in the time interval $[t,t+\Delta t]$, which occurs with probability $W_{nm}\Delta t$. Similarly, it is equal to $-1$ if $N_i(t)=n$ and $N_i(t+\Delta t)\neq n$, that is, there is a transition $n\rightarrow m$ for some $m\neq n$ with probability $W_{mn}\Delta t$.}we find that
\begin{eqnarray}
\fl &\sum_{n=0}^{K-1}\int_{\R^d}d\x\,f_n(\x)\frac{\partial \rho_i(\x,n,t)}{\partial t} 
=\sum_{n=0}^{K-1}\int_{\R^d}d\x\, \rho_i(\x,n,t)\bigg [\sqrt{2D}{\bm \nabla} f_n(\x)\cdot {\bm \xi}_i(t)+D{\bm \nabla}^2 f_n(\x)\nonumber \\
\fl &\qquad +{\bm \nabla} f_n(\x)\cdot {\bf A}_n(\x)\bigg ] +  \sum_{m=0}^{K-1}Q_{mn}\int_{\R^d}d\x\, \rho_i(\x,m,t) f_n(\x).
\end{eqnarray}
Integrating by parts various terms on the right-hand side, and exploiting the arbitrariness of the functions $f_n$ yields the following hSPDE:
\begin{eqnarray}
\label{stage}
\fl \frac{\partial \rho_i(\x,n,t)}{\partial t} 
 &= \int_{\R^d}d\x\, \bigg \{-\sqrt{2D} {\bm \nabla} \cdot   [ \rho_i(\x,n,t) {\bm \xi}_i(t) +D{\bm \nabla}^2  \rho_i(\x,n,t) \\
 \fl & -{\bm \nabla} \cdot [ \rho_i(\x,n,t){\bf A}_n(\x)]\bigg \} +   \sum_{m=0}^{K-1}Q_{nm}\rho_i(\x,m,t) .\nonumber 
\end{eqnarray}
Finally, 
summing over the particle index $i$ and using the definition of the global densities ${\rho}_n(\x,t)$ gives
\begin{eqnarray}
\fl \frac{\partial  {\rho}_n(\x,t)}{\partial t} 
 &=-\sqrt{2D}\sum_{i=1}^{\calN}{\bm \nabla} \cdot \bigg [ \rho_i(\x,n,t) {\bm \xi}_i(t)\bigg ]+D{\bm \nabla}^2  \rho_n(\x,t)-{\bm \nabla} \cdot [ \rho_n(\x,t){\bf A}_n(\x)]\nonumber \\
 \fl &\quad +\sum_{m=0}^{K-1}Q_{nm}\rho_m(\x,t).
\label{drho1}
\end{eqnarray}

\begin{figure}[b!]
 \centering
\includegraphics[width=12cm]{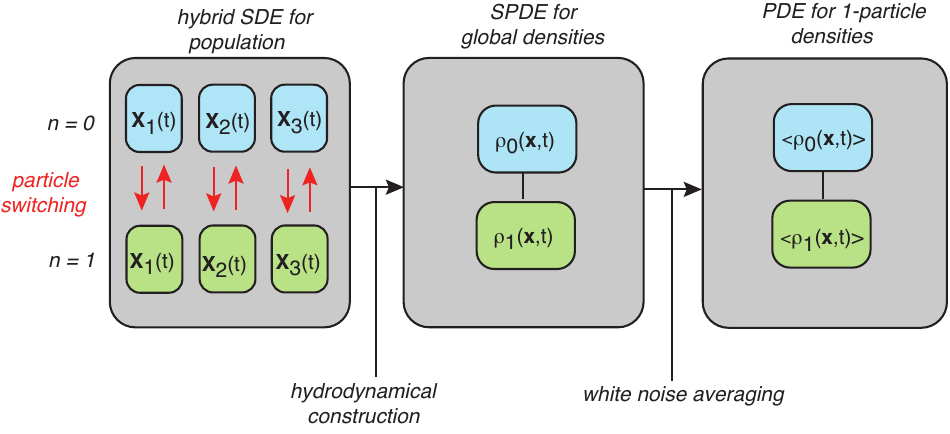}
\caption{Corresponding hierarchy of models for the global density functions of a population of non-interacting particles that independently switch between different internal states.}\label{fig3}
\end{figure}

As in the analysis of environmental switching, see equation (\ref{rho1}), we do not have a closed equation for $\rho_n$.  However, we can generalize the construction of Ref. \cite{Dean96} by explicitly taking into account the discrete index $n$. More specifically, we introduce the space-dependent Gaussian noise terms
\begin{equation}
{\xi}_n(\x,t)=-\sum_{i=1}^{\calN}{\bm \nabla} \cdot \bigg [ \rho_i(\x,n,t) {\bm \xi}_i(t)\bigg ]
\end{equation}
with zero mean and the correlation function
\begin{equation}
\fl \langle 
{\xi}_n(\x,t){\xi}_m(\y,t')\rangle = \delta(t-t')\sum_{i=1}^{\calN} {\bm \nabla}_{\x}\cdot {\bm \nabla}_{\y}\bigg (\rho_i(\x,n,t) \rho_i(\y,m,t) \bigg ).
\end{equation}
Since 
\begin{eqnarray}
\fl \rho_i(\x,n,t) \rho_i(\y,m,t) &=\delta(\x-\X_i(t))\delta(\y-\X_i(t))\E[\delta_{N_i(t),n}]\E[\delta_{N_i(t),m}]\nonumber \\
&=\delta(\x-\y)\delta_{n,m}\rho_i(\y,n,t),
\end{eqnarray}
 it follows that
\begin{equation}
\fl \langle\xi_n (\x,t)\xi_m(\y,t')\rangle = \delta_{n,m}\delta(t-t')\sum_{i=1}^{\calN} {\bm \nabla}_{\x}\cdot {\bm \nabla}_{\y}\bigg (\delta(\x-\y)\rho_i(\x,n,t) \bigg ).
\end{equation}
Finally, we introduce the global density-dependent noise fields
\begin{equation}
\widehat{\xi }_n(\x,t)={\bm \nabla} \cdot\bigg ({\bm \eta}_n(\x,t)\sqrt{\rho_n(\x,t)}\bigg ),
\end{equation}
where ${\bf \eta}_n(\x,t)$ is a global white noise field whose components satisfy 
\begin{equation}
\langle  \eta_n^{\sigma}(\x,t)\eta_m^{\sigma'}(\y,t')\rangle =\delta_{n,m}\delta(t-t')\delta(\x-\y)\delta_{\sigma,\sigma'}.
\end{equation}
It can be checked that the Gaussian noises ${ \xi}_n$ and $\widehat{\xi}_n$ have the same correlation functions and are thus statistically identical. We thus obtain a closed SPDE for the global densities:
\begin{eqnarray}
\fl \frac{\partial \rho_n(\x,t)}{\partial t} 
&=\sqrt{2D}{\bm \nabla} \cdot \bigg [ \sqrt{\rho_n(\x,t)} {\bm \eta}_n(\x,t)\bigg ]+D{\bm \nabla}^2  \rho_n(\x,t) 
-{\bm \nabla} \cdot  [\rho_n(\x,t){\bf A}_n(\x)] \nonumber \\
\fl &\quad +\sum_{m=0}^{K-1}Q_{nm}\rho_m(\x,t).
\label{drhocav}
\end{eqnarray}
Note that the only source of noise in equation (\ref{drhocav}) is the spatio-temporal noise. Averaging with respect to the latter and setting $u_n(\x,t)=\langle \rho_n(\x,t)\rangle$ yields the deterministic PDE
\begin{eqnarray}
\fl \frac{\partial u_n(\x,t)}{\partial t} 
&=D{\bm \nabla}^2  u_n(\x,t) 
-{\bm \nabla} \cdot  [u_n(\x,t){\bf A}_n(\x)]  +\sum_{m=0}^{K-1}Q_{nm}u_m(\x,t).
\label{drhocav2}
\end{eqnarray}
Equation (\ref{drhocav2}) is identical in form to the CK equation (\ref{CKH}) for a single actively switching process, whereas equation (\ref{drhocav}) is a stochastic version of the CK equation with density-dependent multiplicative noise. The various versions of the population model for intrinsically switching particles are summarized in Fig. \ref{fig3}.

\pcb{\subsection{Quasi-steady-state (QSS) approximation in the weak-noise limit}
As we mentioned in section 2, many switching processes in cell biology involve a separation of time scales between the discrete and continuous processes. A standard method for reducing the complexity of the single-particle hSDE is to use some form of quasi-steady-state (QSS) approximation that reduces the associated Chapman-Kolmogorov (CK) equation to an effective drift-diffusion equation for the marginal probability density $p(\x,t)=\sum_mp_m(\x,t)$ \cite{Papanicolaou75,Reed90,Schnitzer93,Hillen00,Tailleur08,Newby10}. The basic idea is to work on length-scales and time-scales such that the rate of switching is relatively fast. Here we apply the QSS approximation to the stochastic global density equation (\ref{drhocav}), extending the particular formulation of Ref. \cite{Newby10}. For simplicity, we focus on the one-dimensional case.
}

\pcb{Suppose that we rescale the matrix generator ${\bf Q}$ and diffusivity $D$ according to ${\bf Q}\rightarrow {\bf Q}/\epsilon$ and $D\rightarrow \epsilon D$, respectively, where $0<\epsilon \ll 1$ (after an appropriate non-dimensionalization). Taking $\epsilon \rightarrow 0$ then corresponds to the weak-noise limit. 
Decompose the probability densities according to
\begin{equation}
\rho_m(x,t)=\rho(x,t)\sigma_m  +\epsilon w_m(x,t),
\end{equation}
where $\sum_{m} \rho_m(x,t) =\rho(x,t)$ and $\sum_{m} w_m(x,t)=0$. Substituting into equation (\ref{drhocav}) yields
\begin{eqnarray}
\label{1pd2}
\fl &\sigma_n \frac{\partial \rho(x,t)}{\partial t}+\epsilon \frac{\partial w_n(x,t) }{\partial t} \\
\fl &=\sqrt{2\epsilon D}\frac{\partial}{\partial x} \bigg [ \sqrt{[\sigma_n \rho(x,t)+\epsilon w_n(x,t)} { \eta}_n(x,t)\bigg ]+\epsilon D\frac{\partial^2}{\partial x^2} [\sigma_n \rho(x,t)+\epsilon w_n(x,t)]
\nonumber \\
\fl &\quad  -\frac{\partial}{\partial x}\bigg  [ A_n(x)  [\sigma_n \rho(x,t)+\epsilon w_n(x,t) ]\bigg ] + \frac{1}{\epsilon}\sum_mQ_{nm}[\sigma_m \rho(x,t) +\epsilon w_m(x,t)].\nonumber
\end{eqnarray}
Summing both sides with respect to $n$ yields
\begin{eqnarray}
\fl   \frac{\partial \rho(x,t)}{\partial t} &=\sqrt{2\epsilon D}\frac{\partial}{\partial x} \bigg [ \sum_m \sqrt{[\sigma_m \rho(x,t)+\epsilon w_m(x,t)} { \eta}_m(x,t)\bigg ]+\epsilon D\frac{\partial^2}{\partial x^2}  \rho(x,t) 
\nonumber \\
\fl &\quad  -\frac{\partial}{\partial x}  \bigg [\rho(x,t) \overline{A} (x)+\epsilon \sum_mA_m(x)w_m(x,t)\bigg ],  \label{1pd3}\end{eqnarray}
where $\overline{A}(x)=\sum_{m}A_m(x)\sigma_m$. Using (\ref{1pd2}) and the fact that $\sum_{m}Q_{nm}\sigma_m=0$, we find
\begin{eqnarray}
\label{w0}
\fl \epsilon  \frac{\partial w_n(x,t) }{\partial t}&=\sum_mQ_{nm}   w_m(x,t)+\sqrt{2\epsilon D}\frac{\partial}{\partial x} \bigg [ \sqrt{[\sigma_n \rho(x,t)+\epsilon w_n(x,t)} { \eta}_n(x,t)\bigg ]
\nonumber \\
\fl & -\sigma_n\sqrt{2\epsilon D}\frac{\partial}{\partial x}\bigg [    \sum_m \sqrt{[\sigma_m \rho(x,t)+\epsilon w_m(x,t)} { \eta}_m(x,t)\bigg  ]  \nonumber \\
\fl &-\sigma_n \frac{\partial}{\partial x}  [ A_n(x)-\overline{A}(x)] \rho(x,t) \nonumber \\
\fl &+\epsilon \frac{\partial}{\partial x} \bigg [ \sigma_n\sum_mA_m(x)w_m(x,t)-A_n(x)w_n(x,t)\bigg]+\epsilon^2 D\frac{\partial^2 w(x,t)}{\partial x^2}.
\end{eqnarray}
}

\pcb{Equations (\ref{1pd3}) and (\ref{w0}) form a slow/fast dynamical system, which we rewrite in the more compact form
\numparts
\begin{eqnarray}
\label{sfa}
 \fl \partial_t \rho &=-\partial_x[\overline{A}(x)\rho] +\epsilon D\partial_{xx}\rho+\epsilon \sum_m\partial_x [A_m(x)w_m]+\sqrt{2\epsilon D} \sum_m \partial_x[\theta_m^{\epsilon}\eta_m]\\
\fl \partial_t w_n &=\frac{1}{\epsilon}\bigg [\sum_m Q_{nm}w_m-\sigma_n \partial_x  [ A_n(x)-\overline{A}(x)] \rho\bigg ]+\sqrt{\frac{2D}{\epsilon}}  \partial_x[\theta_n^{\epsilon}\eta_n] \nonumber \\
\fl &\quad -\sigma_n\sqrt{\frac{2D}{\epsilon}}  \sum_m \partial_x[\theta_m^{\epsilon}\eta_m], \quad n=0,\ldots,K-1,
\label{sfb}
\end{eqnarray}
\endnumparts
where
\begin{equation}
\theta_m^{\epsilon}(x,t)=\sqrt{\sigma_m \rho(x,t)+\epsilon w_m(x,t)} .
\end{equation}
We have divided equation (\ref{w0}) by $\epsilon$ and dropped the hidher-order diffusion term. Note that $\langle \sum_m\theta_m^{\epsilon}(x,t)\rangle =0$ and
\begin{eqnarray}
\fl & \left \langle \sum_m\theta_m^{\epsilon}(x,t)\eta_m(x,t) \sum_n\theta_n^{\epsilon}(x',t')\eta_n(x',t')\right \rangle \nonumber \\
\fl &= \sum_{m,n}\sqrt{\sigma_m \rho(x,t)+\epsilon w_m(x,t)}\sqrt{\sigma_n \rho(x',t')+\epsilon w_n(x',t')}\langle  { \eta}_m(x,t){ \eta}_n(x',t')\rangle\nonumber \\
\fl &=\sum_n[\sigma_n \rho(x,t)+\epsilon w_n(x,t)]\delta(x-x')\delta(t-t')= \rho(x,t) \delta(x-x')\delta(t-t').
\label{var}
\end{eqnarray}
Since the QSS reduction eliminates the variables $w_n$, we do not need to consider correlations between $\rho$ and $w_n$. Therefore, we can replace the noise term on the right-hand side of equation (\ref{sfa}) by $\sqrt{2\epsilon D}\partial_x[\sqrt{\rho(x,t)} \eta(x,t)]$, where $\eta(x,t)$ is a scalar spatio-temporal white noise process.
 It follows from equations (\ref{sfa}) and (\ref{sfb}) that the dynamics of $w_n$ is an order $1/\epsilon$ fast than $\rho$. The fact that the white noise terms on the right-hand side of equation (\ref{sfb}) live on the fast time-scale follows from the observation that $ \epsilon^{-1/2} \eta_m(\cdot,\epsilon t)dt$ is statistically equivalent to $\eta_m(\cdot, t)dt$.
From the dynamical systems perspective, the fast  dynamics can be restricted to the $(K-1)$-dimensional invariant manifold satisfying the constraint $\sum_{m=0}^{K-1}w_m=0$. We take the coordinates of the invariant manifold to be the components of the vector ${\bf w}=(w_1,\ldots, w_{K-1})$ with $w_0=-\sum_{m=1}^{K-1} w_m$. It follows that, for fixed $x$ and frozen slow variable $\rho(x,t)=\Theta$,  the vector ${\bf w}(x,t)$ evolves according to an SDE parameterized by $\Theta$. Assuming that there exists a unique (ergodic) stationary density $\mu_{\epsilon}({\bf w};\Theta)$ then a slow/fast analysis implies that in the small $\epsilon$ limit, equation (\ref{sfa}) can be written as a closed equation for $\rho$ in which the linear terms involving ${\bf w}$ are averaged with respect to the stationary density $\mu_{\epsilon}$ \cite{Pavliotis08}:
\begin{eqnarray}
\label{sfa2}
 \fl \partial_t \rho &=-\partial_x[\overline{A}(x)\rho] +\epsilon D\partial_{xx}\rho+\epsilon \sum_m\partial_x [A_m(x)\overline{w}_m ]+\sqrt{2\epsilon D} \partial_x[\sqrt{ \rho}\, \eta],
 \end{eqnarray}
 where
 \begin{equation}
\overline{w}_m(x,t) =\int w_m\mu_0({\bf w};\rho(x,t))d{\bf w}.
 \end{equation}
 Finally, averaging the right-hand side of equation (\ref{sfb}) with respect to $\mu_0$ implies that the steady-state first moment satisfies the linear equation
\begin{equation}
\sum_m Q_{nm}\overline{w}_m =\sigma_n \partial_x \bigg [ ( A_n(x)-\overline{A}(x)) \rho \bigg ].
\label{aaa}
\end{equation}
}

\pcb{The Fredholm alternative theorem ensures that equation (\ref{aaa}) has a unique solution on the invariant manifold for fixed $x,t$. Denoting the pseudo-inverse of the matrix ${\bf Q}$ by ${\bf Q}^{\dagger}$, we have
\begin{equation}
\overline{w}_m =\sum_n Q^{\dagger}_{mn}\sigma_n \partial_x  \bigg [ (A_n(x)-\overline{A}(x)) \rho \bigg ].
\end{equation}
Substituting this result back into equation (\ref{sfa}) finally gives a closed SPDE for $\rho$:
\begin{eqnarray}
\label{sfa3}
 \fl \partial_t \rho &=-\partial_x[\overline{A}(x)\rho] +\epsilon D\partial_{xx}\rho+\epsilon \sum_{m,n}\partial_x \bigg [A_m(x)  Q^{\dagger}_{mn}\sigma_n \partial_x  \bigg [ (A_n(x)-\overline{A}(x)) \rho \bigg ]\bigg  ]\nonumber \\
 \fl & \quad +\sqrt{2\epsilon D}  \partial_x[\sqrt{\rho} \eta],
 \end{eqnarray}
 Equation (\ref{sfa3}) can be rewritten in the form
 \begin{eqnarray}
\label{sfa4}
  \partial_t \rho &=-\partial_x[\calV_{\epsilon}(x)\rho] +\epsilon \partial_{xx} [(D+{\mathcal D}(x)])\rho] +\sqrt{2\epsilon D}   \partial_x[\sqrt{ \rho}\, \eta ],
 \end{eqnarray}
 where
 \begin{equation}
 {\mathcal D}(x)=\sum_{m,n}[A_m(x)-\overline{A}(x)] Q^{\dagger}_{mn}\sigma_n   [A_n(x)-\overline{A}(x)],
 \end{equation}
 and
 \begin{equation}
 \calV_{\epsilon}(x)=\overline{A}(x)+\epsilon \sum_{m,n}\partial_x [A_m(x)-\overline{A}(x)] Q^{\dagger}_{mn}\sigma_n   [A_n(x)-\overline{A}(x)].
 \end{equation}
 It follows that in the weak-noise limit, the marginal density $\rho(x,t)$ evolves according to a stochastic Fokker-Planck equation. However, it is not equivalent to the Dean-Kawasaki equation for the global density obtained by applying the QSS reduction to the individual particles first. The latter leads to the independent Ito SDEs
 \begin{equation}
 \fl dX_i=\calV_{\epsilon}(X_i)dt+\sqrt{2\epsilon D}dW_i(t)+\sqrt{2\epsilon {\mathcal D}(x)}d\widehat{W}_i(t),\quad i=1,\ldots ,\calN,
 \end{equation}
 where $W_i(t)$ and $\widehat{W}_i(t)$ are independent Wiener processes. The Dean-Kawasaki equation for the global density is
 \begin{eqnarray}
\fl \frac{\partial \rho(x,t)}{\partial t} 
&= \frac{\partial}{\partial x}\bigg [ \sqrt{2\epsilon D \rho(x,t)}  \eta(x,t)\bigg ]+\frac{\partial}{\partial x}\bigg [ \sqrt{2\epsilon {\mathcal D}(x)\rho(x,t)}  \widehat{\eta}(x,t)\bigg ]\nonumber \\
\fl & \quad +\epsilon \frac{\partial^2} {\partial x^2} [D+{\mathcal D}(x)]\rho(x,t) 
-\frac{\partial}{\partial x} [\rho(x,t)\calV_{\epsilon}(x)] . 
\end{eqnarray}
The white-noise process $\widehat{\eta}$ is the diffusion approximation of the fast switching process. Averaging with respect to $\widehat{\eta}$ recovers equation  (\ref{sfa4}).
 }

\subsection{Interacting Brownian particles and mean field theory}  

\pcb{Further differences between environmental and particle switching arise when particle interactions are included. Let us return to the example of interacting overdamped Brownian particles considered in Sect. 3.3, but now assume that each particle independently switches between different conformational states. Moreover, suppose that the pair-wise interaction between the particles $(\X_j(t),N_j(t))$ and $(\X_k(t),N_k(t))$ is given by $-D\beta{\bm \nabla} K_{N_j(t)N_k(t)} (|\X_j(t)-\X_k(t)|)$. That is, the interaction potential depends on the internal conformational states of the particle pair. (For simplicity, we assume that the effective external potential $V(\x)$ seen by a particle is independent of its internal state.)}
The hybrid SDE of an individual particle takes the form 
\begin{eqnarray}
\fl d\X_j(t)=-D\beta  \pcb{{\bm \nabla}} V(\X_j(t))dt-D\beta\sum_{k=1}^{\calN} \pcb{\bm \nabla}K_{N_j(t)N_k(t)} (|\X_j(t)-\X_k(t)|)+\sqrt{2D}d{\bf W}_j(t)\nonumber \\
\fl
\label{moo}
 \end{eqnarray}
 As before, we introduce the global densities (\ref{dglob}) and follow the various steps used in the derivation of the density equation (\ref{stage}):
 \begin{eqnarray}
\fl \frac{\partial \rho_i(\x,n,t)}{\partial t} 
&=-\sqrt{2D}{\bm \nabla} \cdot [ \rho_i(\x,n,t) {\bm \xi}_i(t)]+D{\bm \nabla}^2  \pcb{\rho_i(\x,n,t)}+D\beta {\bm \nabla} \cdot \pcb{\rho_i(\x,n,t) {\bm \nabla}V(\x)}\nonumber \\
\fl & +\pcb{D\beta \E\bigg [{\bm \nabla} \cdot \widehat{\rho}_i(\x,n,t) \int_{\R^d}  \sum_{j=1}^{\cal N}\sum_{m=0}^{K-1} \widehat{\rho}_j(\y,m,t){\bm \nabla}K_{nm}(\x-\y) d\y\bigg ]}\nonumber \\
\fl &\qquad +\sum_{m=0}^{K-1}Q_{nm}\rho_i(\x,m,t),
\end{eqnarray}
where
\begin{equation}
\widehat{\rho}_i(\x,n,t)=\delta(\X_i(t)-\x)\delta_{N_i(t),n},\quad  {\rho}_i(\x,n,t)=\E[\widehat{\rho}_i(\x,n,t)],
\end{equation}
and expectation is taken with respect to the discrete Markov process.
Summing over the particle index $i$ along identical lines to the non-interacting case then gives
\begin{eqnarray}
\fl \frac{\partial \rho_n(\x,t)}{\partial t} 
&=\sqrt{2D}{\bm \nabla} \cdot \bigg [ \sqrt{\rho_n(\x,t)} {\bm \eta}_n(\x,t)\bigg ]+D{\bm \nabla}^2  \rho_n(\x,t)+\pcb{ \sum_mQ_{nm}\rho_m(\x,t)}\nonumber \\
\fl &\ +D\beta\pcb{\E\bigg [ {\bm \nabla} \cdot  \widehat{\rho}_n(\x,t)\bigg ( {\bm \nabla}V(\x) +\int_{\R^d}  \sum_{m}\widehat{\rho}_m(\y,t) {\bm \nabla}K_{nm}(\x-\y)d\y\bigg )\bigg ]},\label{drhoc2}\end{eqnarray}
with $\widehat{\rho}_n(\x,t)=\sum_i\widehat{\rho}_i(\x,n,t)$. In contrast to the non-interacting case, we no longer obtain a closed SPDE for $ \rho_n(\x,t)$ when taking expectations with respect to the discrete stochastic process. Consequently, averaging with respect to the white noise process yields a PDE for the one-particle density
\begin{equation}
u_n(\x,t)=\langle p_n(\x,t)\rangle =\bigg \langle \E\bigg [\sum_j \delta(\x-\X_j(t))\delta_{N_j(t),n}\bigg ]\bigg \rangle,
\end{equation}
that couples to the two-point correlation function
\begin{equation}
\fl u_{nm}^{(2)}(\x,\y,t)=\bigg \langle \E\bigg [\sum_{j,k} \delta(\x-\X_j(t)) \delta(\y-\X_k(t))\delta_{N_j(t),n}\delta_{N_k(t),m}\bigg ]\bigg \rangle .
\end{equation}
\pcb{That is,
\begin{eqnarray}
\label{drhoc2av}
\fl \frac{\partial u_n(\x,t)}{\partial t} 
&=D{\bm \nabla}^2 u_n(\x,t)+\pcb{ \sum_mQ_{nm}u_m(\x,t)} \\
\fl &\ +D\beta \bigg [ {\bm \nabla} \cdot \bigg ( u_n(\x,t) {\bm \nabla}V(\x) +\int_{\R^d}  \sum_{m}u_{nm}^{(2)}(\x,\y,t) {\bm \nabla}K_{nm}(\x-\y)d\y\bigg )\bigg ].\nonumber 
\end{eqnarray}
This has a completely different structure compared to the corresponding hPDE for and interaction potential $K_n$ that depends on the state of a randomly switching environment. In particular, replacing $A_n(\x)$ and ${\bf F}(\x-\y)$ in equation (\ref{rhocU}) by the terms $-D\beta {\bm \nabla V}(\x)$ and $-D\beta {\bm \nabla}K_n(|\x-\y|)$, respectively, we have
\begin{eqnarray}
\frac{\partial u(\x,t)}{\partial t} &=D{\bm \nabla}^2    u(\x,t) +\beta D{\bm \nabla} \cdot  [u(\x,t) {\bm \nabla}V(\x)]\nonumber \\
&\quad +\beta D{\bm \nabla} \cdot  \int_{\R^d} \langle\rho(\x,t) \rho(\y,t)\rangle {\bm \nabla}K_n(|\x-\y|)d\y  . 
\label{rhocU}
\end{eqnarray}
}

One recent example of an active binary switching system of interacting particles involves a one-component soft colloidal system in which every particle can individually stochastically switch between two interaction states \cite{Bley21,Bley22}. The two states correspond to a `small' ($n=0$) and `big' ($n=1$) conformational state, respectively, such that the interaction potential is given by an indexed Gaussian:
\begin{equation}
K_{nm}(x)=a_{nm}\e^{-x^2/\sigma_{nm}^2},\quad n,m=0,1.
\end{equation}
It is also assumed that the external potential does not switch.  The discrete state $N(t)\in \{0,1\}$ evolves according to a two-state Markov chain with matrix generator 
 \begin{equation}
 {\bf Q}=\left (\begin{array}{cc} -\gamma  &\alpha  \\ \gamma & -\alpha  \end{array} \right )
 \label{2M}
 \end{equation}
\pcb{ In Refs. \cite{Bley21,Bley22} it is assumed that the mean field limit holds in the case of weakly interacting switching particles, which then leads to a closed system of equations for $u_n(\x,t)$:
\numparts
\begin{eqnarray}
\label{DDFTa}
\frac{\partial u_0(\x,t)}{\partial t}=-{\bm \nabla}\cdot {\bf J}_0(\x,t)-\gamma u_0(\x,t)+\alpha u_1(\x,t),\\
\frac{\partial u_1(\x,t)}{\partial t}=-{\bm \nabla}\cdot {\bf J}_1(\x,t)+\gamma u_0(\x,t)-\alpha u_1(\x,t),
\label{DDFTb}
\end{eqnarray}
\endnumparts
where
\begin{eqnarray}
\fl {\bf J}_n(\x,t)=-D {\bm \nabla} u_n(\x,t)-\beta D u_n(\x,t){\bm \nabla} \bigg [V(\x)+ \sum_{m=0}^{K-1}\int_{\R^d}K_{nm}(|\x-\y)u_m(\y,t)d\y\bigg ].\nonumber \\
\fl 
\label{Janz}
\end{eqnarray}
Finally, as also shown by these authors, substituting equation (\ref{Janz}) into the CK equations (\ref{DDFTa}) and (\ref{DDFTb}), and taking the fast switching limit yields a single equation for the scalar density $u(\x,t)=\rho_0u_0(\x,t)+\rho_1u_1(\x,t)$, where $\sigma_0=\alpha/(\alpha+\gamma)$ and $\sigma_1=\gamma/(\alpha+\gamma)$:
\begin{eqnarray}
\fl \frac{\partial u(\x,t)}{\partial t}=D{\bm \nabla}^2 u(\x,t)+D\beta {\bm \nabla} \cdot 
u(\x,t)\bigg ({\bm \nabla}V(\x)+\int_{\R^d} {\bm \nabla}\overline{K}(\x-\y|) u(\y,t)d\y\bigg ),
\end{eqnarray}
where
\begin{equation}
\overline{K}(|\x-\y|)=\sum_{n,m}\sigma_n\sigma_m K_{nm}(|x-\y|).
\end{equation}
}

\section{Nonequilibrium statistical field theory \pcb{(non-interacting particles)}} The analysis of the stochastic global density equations derived in this paper is nontrivial even in the absence of pair-wise interactions. In the case of a randomly switching environment, the density equation is given by the hSPDE (\ref{rhoc}), whereas for particle switching it takes the form of the SPDE (\ref{drhocav}). One approach is to recast the density equations into a field theory. This provides a framework for performing perturbative series expansions and, in certain cases, yields non-perturbative approximations to various correlation functions.  As a first step in this direction, a field theory for a non-interacting, non-switching Brownian gas has recently been constructed for the global density \cite{Velenich20}. The basic idea is to apply a Martin-Siggia-Rose-Janssen-de Dominicis (MSRJD) path integral construction \cite{Martin73,Dom76,Janssen76} to the Dean equation obtained by setting $A_n(\x)=0$ in equation (\ref{rhoc}). (For a complementary approach based on a Doi-Peliti path integral formulation \cite{Doi76,Doi76a,Peliti85}, see Ref. \cite{Bothe23}. Note, however, that care has to be taken when comparing Doi-Peliti and MSRJD field theories since the actual fields have different physical interpretations.) One of the interesting features of the MSRJD path integral representation is that, even though the particles do not interact, the resulting field theory contains an interaction term. The presence of a 3-vertex reflects the original particle nature of the gas, and ensures that the density field is strictly positive, in contrast to a Gaussian free field. \pcb{In this final section, we indicate how to extend the MSRJD path integral to the more general global density equations obtained in previous sections for non-interacting particle systems. }

\subsection{MSRJD path integral: particle switching} In order to simplify our derivation, we consider a 1D model with two internal states $n=0,1$. Equation (\ref{drhocav}) reduces to the form
\begin{eqnarray}
\fl \frac{\partial \rho_n(x,t)}{\partial t} 
&=\sqrt{2D}\partial_x [ \sqrt{\rho_n(x,t)} { \eta}_n(x,t)\bigg ]+D\partial_x^2  \rho_n(x,t) 
-\partial_x  [\rho_n(x,t){A}_n(x)] \nonumber \\
\fl &\quad +\sum_{m=0,1}Q_{nm}\rho_m(x,t),
\label{DL}
\end{eqnarray}
with the matrix generator given by equation (\ref{2M}). The first step in the MSRJD procedure is to discretize equation (\ref{DL}) by dividing the time interval $[0,t]$ into $M$ equal subintervals of size $\Delta t$ and setting
\numparts
\begin{eqnarray}
\phi_{\ell}(x)&=\P[X_{\ell}=x,N_{\ell}=0]=\rho_0(x,\ell \Delta t),\\ \psi_{\ell}(x)&=\P[X_{\ell}=x,N_{\ell}=1]=\rho_1(x,\ell \Delta t).
\end{eqnarray}
\endnumparts
with fixed initial densities $\phi_0(x)$ and $\psi_0(x)$. Equation (\ref{DL}) becomes
\numparts
\begin{eqnarray}
\label{ellphi}
\fl  \phi_{\ell+1}(x)&=\phi_{\ell}(x)+\bigg[\L_0\phi_{\ell}(x)+\alpha \psi_{\ell}(x)-\gamma \phi_{\ell}(x) \bigg]\Delta t+\sqrt{2D}\frac{d\sqrt{\phi_{\ell}(x)}\Delta W_{0,\ell}(x)}{dx}, \\
 \label{ellphi2}
  \fl \psi_{\ell+1}(x)&=\psi_{\ell}(x)+\bigg [\L_1 \psi_{\ell}(x) -\alpha \psi_{\ell}(x)+ \gamma \phi_{\ell}(x)\bigg ]\Delta t +\sqrt{2D}\frac{d\sqrt{\psi_{\ell}(x)}\Delta W_{1,\ell}(x)}{dx},
\end{eqnarray}
\endnumparts
with $\ell=0,\ldots,M-1$, and $\L_n$ are the linear operators
\begin{equation}
 \label{Lenny}
 \L_n f(x)=- \frac{d (A_n(x)f(x)}{d x}+D\frac{d^2f(x)}{dx^2}.
 \end{equation}
 Moreover $\Delta W_{n,\ell}(x)$ is a Gaussian random variable with zero mean and two-point correlation
 \begin{equation}
 \langle \Delta W_{n,\ell}(x) \Delta W_{n',\ell'}(y)\rangle =\delta_{\ell,\ell'}\delta_{n,n'} \delta(x-y)\Delta t.
 \end{equation}

Consider a particular realization of the spatiotemporal Gaussian noise processes, which we represent by the symbol $\Omega$. Defining the vectors ${\Phi}=(\phi_1,\ldots,\phi_M)$ and $\Psi=(\psi_1,\ldots,\psi_M)$, we introduce the conditional probability density functional 
\begin{eqnarray}
\fl  & {\mathcal P}[{\Phi},\Psi|\phi_0,\psi_0,\Omega]\\
\fl &= \prod_{\ell=0}^{M-1} \prod_x \delta \bigg (\phi_{\ell+1}(x)-\phi_{\ell}(x) -\left [L_0\phi_{\ell}(x)+  \alpha \psi_{\ell}(x)-\gamma \phi_{\ell}(x)\right ]\Delta t -\sqrt{2D} \Delta \widehat{W}_{0,\ell}(x) \bigg )\nonumber \\
 \fl &\times  \prod_{\ell=0}^{M-1} \prod_x \delta \bigg (\psi_{\ell+1}(x)-\psi_{\ell}(x) -\left [\L_1 \psi_{\ell}(x)-  \alpha \psi_{\ell}(x)+\gamma\phi_{\ell}(x)\right ]\Delta t-\sqrt{2D} \Delta \widehat{W}_{1,\ell}(x)  \bigg ).\nonumber 
 \end{eqnarray}
 We have used the compact notation
 \begin{equation}
  \fl \Delta \widehat{W}_{0,\ell}(x) =\frac{d\sqrt{\phi_{\ell}(x)}\Delta W_{0,\ell}(x)}{dx},\quad \Delta \widehat{W}_{1,\ell}(x) =\frac{d\sqrt{\psi_{\ell}(x)}\Delta W_{1,\ell}(x)}{dx}.
  \end{equation}
  Introducing Fourier representations of the Dirac delta functions gives
 \begin{eqnarray}
\fl & {\mathcal P}[{\Phi},\Psi|\phi_0,\psi_0,\Omega]=\int D[\widetilde{\Phi},\widetilde{\Psi}] \\
\fl &\quad \times\exp\bigg \{i\sum_{\ell=0}^{M-1} \int dx\, \widetilde{\phi}_{\ell+1}(x)\bigg [\phi_{\ell+1}(x)-\phi_{\ell}(x)-\bigg (\L_0\phi_{\ell}(x)+\alpha \psi_{\ell}(x)-\gamma \phi_{\ell}(x)  \bigg )\Delta t  \bigg ]\bigg \}\nonumber \\
 \fl& \quad \times \exp\bigg \{i\sum_{\ell=0}^{M-1} \int \widetilde{\psi}_{\ell+1}(x)\bigg [\psi_{\ell+1}(x)-\psi_{\ell}(x)- \bigg (\L_1 \psi_{\ell}(x)-\alpha \psi_{\ell}(x)+\gamma \phi_{\ell}(x) \bigg )\Delta t \bigg ]\bigg \}\nonumber\\
 \fl &\quad \times \exp\bigg \{\sqrt{2D} \sum_{\ell=0}^{M-1} \int dx\, \bigg [\partial_x[i\widetilde{\phi}_{\ell}(x)]\sqrt{\phi_{\ell}(x)}\Delta {W}_{0,\ell}(x)\nonumber \\
 \fl &\hspace{2cm} +\partial_x[i\widetilde{\psi}_{\ell+1}(x)]\sqrt{\psi_{\ell+1}(x)}\Delta  {W}_{1,\ell}(x)\bigg ]\bigg \},
\end{eqnarray}
where $D[\widetilde{\Phi},\widetilde{\Psi}]=\prod_{\ell=0}^{M-1}\prod_x d\widetilde{\phi}_{\ell}(x)d\widetilde{\psi}_{\ell}(x)$. 
We have integrated by parts in the final exponential factor. The final steps are to integrate with respect to the Gaussian processes and then to take the continuum limit $\Delta t \rightarrow 0$ and $M\rightarrow \infty$ with $M\Delta t =t$ and $\widetilde{\phi}(x,\ell \Delta t)=\widetilde{\phi}_{\ell}(x)$ etc. After performing the Wick rotation $(\widetilde{\phi},\widetilde{\psi})\rightarrow (i\widetilde{\phi},i\widetilde{\psi})$ we obtain a formal path integral representation of the probability density functional
\begin{eqnarray}
\label{gen}
{\mathcal P}[\phi,\psi]&=\int{\mathcal D}[\widetilde{\phi},\widetilde{\psi}]\exp\left (-S[\phi,\widetilde{\phi},\psi,\widetilde{\psi}]\right )
 \end{eqnarray}
where 
\begin{eqnarray}
 \fl &S[\phi,\widetilde{\phi},\psi,\widetilde{\psi}]\nonumber \\
 \fl &=\int_0^td\tau \int_{-\infty}^{\infty}dx\, \bigg \{\widetilde{\phi}(x,\tau)\bigg [\partial_{\tau} \phi(x,\tau)+\gamma \phi(x,\tau)+\partial_x[A_0(x)\phi(x,\tau)]-D\partial_{xx}\phi(x,\tau)\bigg ]\nonumber\\
\fl &\quad+\widetilde{\psi}(x,\tau)\bigg [\partial_{\tau} \psi(x,\tau)+\alpha \psi(x,\tau)+\partial_x[A_1(x)\psi(x,\tau)]-D\partial_{xx}\psi(x,\tau) \bigg ]
\nonumber \\
\fl &\quad -\bigg[\alpha \widetilde{\phi}(x,\tau)\psi(x,\tau)+\gamma\widetilde{\psi}(x,\tau)   \phi(x,\tau) \bigg ] \nonumber \\
\fl & \quad -D\bigg [\phi(x,\tau)(\partial_x\widetilde{\phi}(x,\tau))^2+\psi(x,\tau)(\partial_x\widetilde{\psi}(x,\tau))^2\bigg ]\bigg \}+
S_{\rm IC}[\phi,\widetilde{\phi},\psi,\widetilde{\psi}]. 
\label{1act}
\end{eqnarray}
We have incorporated the initial conditions into the path integral by adding the following terms to the action (\ref{1act}):
\begin{eqnarray}
\fl S_{IC}=\int_0^td\tau \int_{-\infty}^{\infty}dx\, \bigg \{\widetilde{\phi}(x,\tau)\delta(\tau)  [\phi(x,\tau)-\rho_0(x)]+\widetilde{\psi}(x,\tau)\delta(\tau) [\psi(x,\tau)-\rho_1(x)]\bigg\}.\nonumber \\
\fl
\end{eqnarray}

\subsection{Moment generating functional}

One typically uses path integrals to calculate expectations of the various fields and their composites. In particular, important quantities such as two-point correlations can be obtained by taking functional derivatives of the moment generating functional
\begin{eqnarray}
\label{GF}
\fl & Z[{\bf h},{\bf \h}] \\
\fl  &=\int{\mathcal D}[\phi,\widetilde{\phi},\psi,\widetilde{\psi}] \exp\bigg(-S[\phi,\widetilde{\phi},\psi,\widetilde{\psi}]+\{ \widetilde{\phi},h_0\} +\{\widetilde{\psi},h_1\}+\{  {\phi},\h_0\}+\{ {\psi},\h_1\} \bigg ),\nonumber \end{eqnarray} 
where ${\bf h}=(h_0,h_1)$, ${\bf \h}=(\h_0,\h_1)$ and
\begin{equation}
\{ a,b\} :=\int_{0}^td\tau \int_{-\infty}^{\infty}dx\, a(x,\tau)b(x,\tau).
\end{equation}
For example,
\begin{equation}
\langle \phi(x,t)\phi(y,t_0)\rangle =\frac{1}{Z}\left .\frac{\delta^2 Z[{\bf h},{\bf \h}] }{\delta \widetilde{h}_0(x,t)\delta \widetilde{h}_0(y,t_0)}\right |_{{\bf h}=0=\widetilde{\bf h}}
\end{equation}
etc. (Note that the normalization of the path integral measure cancels in the definition of expectations such as $
\langle \phi(x,t)\phi(y,t_0)\rangle$.)
Suppose that we decompose the action (\ref{1act}) according to 
\begin{eqnarray}
\fl & S[\phi,\widetilde{\phi},\psi,\widetilde{\psi}]=\{ \widetilde{\phi}, (\partial_{\tau}+\gamma -\L_0)\phi\}  +\{ \widetilde{\psi} ,(\partial_{\tau}+\alpha-\L_1)\psi\}- S_I[\phi,\widetilde{\phi},\psi,\widetilde{\psi}],
 \label{action}
\end{eqnarray}
with $\L_{n}$ given by equation (\ref{Lenny}) and
\begin{equation}
\fl S_I[\phi,\widetilde{\phi},\psi,\widetilde{\psi}]= \alpha \{ \widetilde{\phi} ,\psi\}+\gamma\{\widetilde{\psi} , \phi   \}+D\bigg [\{\phi,(\partial_x\widetilde{\phi})^2\}+\{\psi ,(\partial_x\widetilde{\psi})^2\}\bigg ]-S_{\rm IC}[\phi,\widetilde{\phi},\psi,\widetilde{\psi}] .
\end{equation}
The generating functional (\ref{GF}) can then be rewritten as
\begin{eqnarray}
 \fl Z[{\bf h},{\bf \h}]&=\exp \left ( S_I(\delta/\delta \h_0,\delta/\delta h_0,\delta/\delta \h_1,\delta/\delta h_1)\right ) \ Z_0[h_0,\h_0]Z_1[h_1,\h_1],
 \label{Zoo}
\end{eqnarray}
where
\numparts
\begin{eqnarray}
\fl   Z_0[h_0,\h_0]&=\int{\mathcal D}[\phi,\widetilde{\phi}] \exp\bigg (-\{\widetilde{\phi}, (\partial_{\tau}+\gamma-\L_0)\phi\} +\{ \widetilde{\phi},h_0\}+\{ {\phi},\h_0\} \bigg ), 
 \label{Z0}\\
  \fl Z_1[h_1,\h_1]& 
  =\int{\mathcal D}[\psi,\widetilde{\psi}] \exp\bigg (-\{ \widetilde{\psi}, (\partial_{\tau}+\alpha-\L_1)\psi\}+\{\widetilde{\psi},h_1\}+\{  {\psi},\h_1\} \bigg ). 
  \label{Z1}
\end{eqnarray} 
\endnumparts
Evaluating the Gaussian integrals (\ref{Z0}) and (\ref{Z1}) gives
\begin{eqnarray}
 \fl Z_n[h_n,\h_n]=\exp\left (\int d\tau d\tau'  dx dx'\,  \h_n(x,\tau) G_n(x,\tau|x',\tau') h_n(x',\tau')  \right ), 
  \end{eqnarray}
 where $G_{n}(x,t|x_0,t_0)$ are casual Green's functions. That is,
 \numparts
 \begin{eqnarray}
 \label{Gee}
 G_{0}(x,t|x_0,t_0)&=\e^{-\gamma(t-t_0)} p_{0}(x,t|x_0,t_0)H(t-t_0),\\
 \label{Gee2}
  G_{1}(x,t|x_0,t_0)&=\e^{-\alpha(t-t_0)} p_{1}(x,t|x_0,t_0)H(t-t_0),
  \end{eqnarray}
  \endnumparts
 with $p_{n}(x,t|x_0,t_0)$ the solution to the Fokker-Planck equation
 \begin{equation}
 \label{Lpm}
\frac{\partial p_n}{\partial t} =\L_n p_n \equiv-\frac{\partial [A_n(x)p_n]}{\partial x}+D\frac{\partial^2p_n}{\partial x^2},
\end{equation}
under the initial condition $p_{n}(x,t_0|x_0,t_0)=\delta(x-x_0)$. The exponential factors $\e^{-\gamma(t-t_0)} $ and $\e^{-\alpha(t-t_0)}$ appearing in equations (\ref{Gee}) and (\ref{Gee2}) are the probabilities that there are no transitions $0\rightarrow 1$ and $1\rightarrow 0$, respectively, over the time interval $[t_0,t]$. 

One non-trivial example for which $p_n$ can be calculated explicitly is an OU process with random drift.
 This particular  hSDE has been used to model an RTP with diffusion in a harmonic potential \cite{Basu20,Garcia21} and protein synthesis in a gene network \cite{Bose04,Smiley10}. In the former case, $X(t)\in \R$ represents the position of the RTP at time $t$ whereas $N(t)=n\in \{0,1\}$ specifies the current velocity state $v_n$ of the particle. If $v_0=v$ and $v_1=-v$ then the motion becomes unbiased when the mean time spent in each velocity state is the same \pcb{($\alpha=\gamma$)}. On the other hand, in the case of the gene network, $X(t)$ represents the current concentration of synthesized protein and $N(t)$ specifies whether the gene is active or inactive. That is, $v_n$ is the rate of synthesis with $v_0>v_1\geq 0$. In both examples, the variable $X(t)$ evolves according to the hSDE
\begin{equation}
\label{PDMPrtp}
dX(t)=[-\kappa_0X(t)+v_n]dt+\sqrt{2D}dW(t),\quad N(t)=n,
 \end{equation}
 where $\kappa_0$ represents an effective ``spring constant'' for an RTP in a harmonic potential, whereas it corresponds to a protein degradation rate in the case of a gene network. Comparison with equation (\ref{PDMP}) implies that $A_n(x)=-\kappa_0 x+v_n$. One major difference between an RTP and a gene network is that the continuous variable $X(t)$ has to be positive in the latter case. However, one often assumes that the effective ``harmonic potential'' for $v_0>v_1\geq 0$ restricts $X(t)$ to positive values with high probability so that the condition $X(t)\geq 0$ does not have to be imposed. explicitly. (If $D=0$ then $X(t)\in \Sigma = [v_0/\kappa_0,v_1/\kappa_0]$ and the CK equation can be restricted to the finite interval $\Sigma$ with reflecting boundary conditions at the ends. In this case, the steady-state CK equation can be solved explicitly \cite{Kepler01,Bose04,Smiley10}.) In the case of an OU process with random drift, one finds that
\begin{eqnarray}
\label{pdfOU2}
\fl & p_{n}(x,t|x_0,0) =\frac{1}{\sqrt{2\pi \Sigma(t)}}\exp\left (-\frac{[x-x_0\e^{-\kappa_0t}-v_n(1-\e^{-\kappa_0t})/\kappa_0]^2}{2\Sigma(t)}\right ), 
\end{eqnarray}
with 
\begin{equation}
\Sigma(t)=\frac{D}{\kappa_0}(1-\e^{-2\kappa_0t}).
\end{equation}

Equation (\ref{Zoo}) is the starting point for performing various diagrammatic expansions by Taylor expanding the functional operator $\e^{S_I}$. This has been carried out elsewhere for a non-switching Brownian gas \cite{Velenich20,Bothe23} whose MSRJD action is of the form
\begin{eqnarray}
 S[\phi,\widetilde{\phi}]=\{ \widetilde{\phi}, (\partial_{\tau}-D\partial_x^2 )\phi\} -D \{\phi,(\partial_x\widetilde{\phi})^2\}+ S_{\rm IC}[\phi,\widetilde{\phi}].
\end{eqnarray}
The cubic term on the right-hand side generates 3-vertices in any diagrammatic expansion of the path integral, and these play a key role in ensuring positivity of the global density. On the other hand, suppose that we first average the global density equation (\ref{DL}) with respect to the spatiotemporal white noise. This yields a deterministic PDE for the first moments $u_n(x,t)=\langle \rho_n(x,t)\rangle $ given by the CK equation
\numparts
\begin{eqnarray}
\frac{\partial u_0}{\partial t}=\L_0u_0(x,t)-\gamma u_0(x,t) +\alpha u_1(x,t),\\
\frac{\partial u_1}{\partial t}=\L_1u_1(x,t)+\gamma u_0(x,t) -\alpha u_1(x,t)
\end{eqnarray}
\endnumparts
Although this is a deterministic system, it is still possible to carry out the MSRJD construction to obtain the generating functional (\ref{GF}) with the action functional  (\ref{action}) such that $S_I= \alpha \{ \widetilde{\phi} ,\psi\}+\gamma\{\widetilde{\psi} , \phi   \}+S_{\rm IC}$. An expansion of $\e^{S_I}$ now generates contributions involving a fixed number of switching events. 
(Note that a Doi-Peliti version of this path integral construction has recently been applied to the particular example of a single RTP with diffusion in a 1D harmonic potential \cite{Garcia21}. Analogous path integrals have also been developed for RTPs in higher dimensions \cite{Zhang22} and active Ornstein-Uhlenbeck (OU) particles \cite{Bothe21}.)

\subsection{MSRJD path integral: environmental switching} Developing a corresponding MSRJD field theory for environmental switching is more involved. For the sake of illustration, consider a 1D, 2-state version of the hSPDE (\ref{rhoc}). As a further. simplification, we average with respect to the white noise to obtain the hPDE
\begin{eqnarray}
\label{FTpc}
&\frac{\partial u(x,t)}{\partial t} =D\frac{\partial^2  u(x,t)}{\partial x^2}-\frac{\partial  ( u(x,t)A_n(x)}{\partial x} ,\quad N(t)=n,
\end{eqnarray}
with $N(t)$ switching according to a 2-state Markov chain with matrix generator (\ref{2M}). Following along analogous lines to the construction of section 5.1, we discretize equation (\ref{FTpc}) by dividing the time interval $[0,t]$ into $M$ equal subintervals of size $\Delta t$ and setting $u_{\ell}(x)=u(x,\ell \Delta t)$
with a fixed initial density $u_0(x)$ that is twice differentiable. Equation (\ref{FTpc}) becomes
\begin{eqnarray}
\label{ellphiFT}
u_{\ell+1}(x)&=u_{\ell}(x)+ \L_nu_{\ell}(x)\Delta t,\quad N(t)=n\end{eqnarray}
with $\ell=0,\ldots,M-1$, and $\L_n$ defined in equation (\ref{Lenny}). 
Consider a particular realization of the discrete stochastic process  $N(\ell\Delta t)=n_{\ell}$ and set ${\bf n}=(n_0,\ldots,n_{M-1})$. Defining $U=(u_1,\ldots,u_M)$, we introduce the conditional probability density functional 
\begin{eqnarray}
 {\mathcal P}[U|u_0,{\bf n}]= \prod_{\ell=0}^{M-1} \prod_x \delta \bigg (u_{\ell+1}(x)-u_{\ell}(x) - L_{n_{\ell}}u_{\ell}(x)\Delta t\bigg ) .\end{eqnarray}
Inserting the Fourier representation of the Dirac delta function gives 
\begin{eqnarray} 
\fl {\mathcal P}(U|u_0,{\bf n}) =\int D[\widetilde{U}]\exp\bigg \{ i\sum_{\ell=0}^{M-1}\int dx\, \widetilde{u}_{\ell+1}(x)[u_{\ell+1}(x)-u_{\ell}(x) -\L_{n_{\ell}}u_{\ell}(x)\Delta t
\bigg \}
\end{eqnarray}
If we now average over the intermediate discrete states $n_{\ell},\ell=1,M-1$ then
\begin{eqnarray}
\fl {\mathcal P}(U,n_M|u_0,n_0)&=\int D[\widetilde{U}]\exp\bigg \{ i\sum_{\ell=0}^{M-1}\int dx\, \widetilde{u}_{\ell+1}(x)[u_{\ell+1}(x)-u_{\ell}(x) ]
\bigg \}\nonumber \\
 \fl &\quad \times \left [\e^{[{\bf Q}+{\bf K}[\widetilde{u}_M,u_{M-1}]\Delta t}\cdots \e^{[{\bf Q}+{\bf K}[\widetilde{u}_1,u_{0}]\Delta t}\right ]_{n_Mn_0},
 \label{discsh}
 \end{eqnarray}
where 
\begin{equation}
\fl {\bf K}[\widetilde{u}_{\ell+1},u_{\ell}]= \left (\begin{array}{cc} -i\int dx\, \widetilde{u}_{\ell+1}(x)\L_{0}u_{\ell}(x)& 0 \\ 0 &-i\int dx\, \widetilde{u}_{\ell+1}(x)\L_{1}u_{\ell}(x)\end{array}\right ).
\end{equation}

In order to obtain a meaningful action functional in the continuum limit, it is necessary to diagonalize the matrix products on the second line of equation (\ref{discsh}). One method is to use coherent spin states along analogous lines to the study of hSDEs for gene networks \cite{Sasai03,Zhang13,Bhatt20,Bressloff21b}. This requires the introduction of auxiliary variables for the path integral action. For a two-state hybrid systems, we first decompose the matrix ${\bf H}={\bf K}+{\bf Q}$ using the Pauli spin matrices
\begin{equation}
\fl \sigma_x=\frac{1}{2}\left (\begin{array}{cc} 0&1 \\ 1& 0 \end{array} \right ),\quad \sigma_y=\frac{1}{2}\left (\begin{array}{cc} 0&-i \\ i& 0 \end{array} \right ),
\quad \sigma_z=\frac{1}{2}\left (\begin{array}{cc} 1&0 \\ 0& -1 \end{array} \right ),
\end{equation}
That is
\begin{eqnarray}
\label{sh:H}
\fl  {\bf H}
 &= \left (\frac{1}{2}{\mathbf 1}  +\sigma_z\right) {K}_0+\left (\frac{1}{2}{\mathbf 1} -\sigma_z\right)  {K}_1-\gamma \left (\frac{1}{2}{\mathbf 1} +\sigma_z\right) -\alpha \left (\frac{1}{2}{\mathbf 1} -\sigma_z\right) +\alpha \sigma_++\gamma\sigma_-,\nonumber
\end{eqnarray}
where $\sigma_{\pm}=\sigma_x\pm i\sigma_y$ and $K_n[u,\widetilde{u}=$. Next we define the coherent spin-$1/2$ state \cite{Radcliffe71}
\begin{equation}
|s\rangle = \left (\begin{array}{c} \e^{i\phi/2}\cos^2\theta/2  \\ \e^{-i\phi/2}\sin^2\theta/2 \end{array} \right ),\quad 0\leq \theta \leq \pi,\ 0\leq \phi <2\pi,
\end{equation}
together with the adjoint
\begin{equation}
\langle s|=\left ( \e^{-i\phi/2}  ,\,  \e^{i\phi/2}  \right ).
\end{equation}
Note that 
\begin{equation}
\langle s'|s\rangle =\e^{i(\phi-\phi')/2}\cos^2\theta/2+\e^{-i(\phi-\phi')/2}\sin^2\theta/2,
\end{equation}
so that $\langle s|s\rangle =1$ and
\begin{equation}
\label{sdiff}
\langle s+\Delta s |s\rangle =1-\frac{1}{2}i \Delta \phi \cos \theta +O(\Delta \phi^2).
\end{equation}
We also have the completeness relation
\begin{equation}
\label{comp2}
\frac{1}{2\pi} \int_0^{\pi}\sin \theta \, d\theta \int_0^{2\pi}d\phi\, |s\rangle \langle s|=1.
\end{equation}
It can checked that the following identities hold:
\begin{eqnarray}
\fl \langle s|\sigma_z|s\rangle =\frac{1}{2}\cos \theta, \
\langle s|\sigma_+|s\rangle = \frac{1}{2}\e^{i\phi}\sin\theta, \
\langle s|\sigma_-|s\rangle =  \frac{1}{2}\e^{-i\phi}\sin\theta .
\end{eqnarray}
Hence,
\begin{eqnarray}
\label{HH}
\fl \langle s|{\bf H}|s\rangle&=H[\theta,\phi,u,\widetilde{u}]\\
\fl &\equiv  -\left (\gamma \left [1-\e^{i\phi}\right ]-K_0[u,\widetilde{u}]\right )\frac{1+\cos\theta}{2} - \left (\alpha \left [1-\e^{-i\phi}\right ]-K_1[u,\widetilde{u}]\right )\frac{1-\cos\theta}{2} .\nonumber 
\end{eqnarray}

We can now diagonalize the matrix product on the second line of equation. (\ref{discsh}) by inserting multiple copies of the completeness relations (\ref{comp2}). Introducing the solid angle integral
\begin{equation}
\int_{\Omega}ds =\frac{1}{2\pi} \int_0^{\pi}\sin \theta \, d\theta \int_0^{2\pi}d\phi,
\end{equation}
 we have
\begin{eqnarray}
\fl & \int_{\Omega}ds_0 \cdots \int_{\Omega}ds_M  
|s_M \rangle \langle s_M|\e^{{\bf H}[\widetilde{u}_M,u_{M-1}]\Delta t}|s_{N-1} \rangle \langle s_{N-1} |\e^{ {\bf H}[\widetilde{u}_{M-1}u_{M-2}]\Delta t}|s_{N-2}\rangle\nonumber \\
\fl  &\cdots \times \langle s_1|\e^{{\bf H}[\widetilde{u}_1u_0]\Delta t}|s_0\rangle \langle s_0|{\bm \psi}(0)\rangle .
 \label{pip}
\end{eqnarray}
In the limit $M\rightarrow \infty$ and $\Delta t \rightarrow 0$ with $M\Delta t =t$ fixed, we can make the approximation
\begin{eqnarray}
\fl \langle s_{\ell+1} |\e^{\widehat{\bf H}\Delta t}| s_{\ell}\rangle  =\langle s_{\ell+1} | s_{\ell} \rangle\bigg \{1+ H[\theta_{\ell},\phi_{\ell},u_{\ell},\widetilde{u}_{\ell+1}]\Delta t \bigg \}+O(\Delta t^2),
\end{eqnarray}
with $H$ defined in equation (\ref{HH}).
In addition, equation (\ref{sdiff}) and the restriction to continuous paths in the continuum limit implies that
\begin{eqnarray}
\fl \langle s_{\ell+1}|s_{\ell}\rangle = 1-\frac{1}{2}i(\phi_{\ell+1}-\phi_{\ell})\cos \theta_{\ell}+O(\Delta \phi^2)=1-\frac{1}{2}i\Delta t\frac{d\phi_{\ell}}{dt}\cos\theta_{\ell} +O(\Delta t^2).
\end{eqnarray}
Hence,
\begin{eqnarray}
\label{K1}
 &\langle s_{\ell+1}|\e^{{\bf H}\Delta t}| s_{\ell}\rangle \approx \exp\left (\left [H[\theta_{\ell},\phi_{\ell},u_{\ell},\widetilde{u}_{\ell+1}]-\frac{i}{2}\frac{d\phi_{\ell}}{dt}\cos\theta_{\ell} \right ]\Delta t\right ) .
 \end{eqnarray}
We can now take the continuum limit.
After Wick ordering, integrating by parts the term involving $d\phi/dt$, and performing the change of coordinates $z=(1+\cos \theta)/2$, we obtain the following functional path integral:
\begin{eqnarray}
\label{gen}
{\mathcal P}_{nn_0}[u]&=\int{\mathcal D}[\theta]{\mathcal D}[z]  {\mathcal D}[\widetilde{u}]\exp\left (-S[u,\widetilde{u},z,\theta]\right )
 \end{eqnarray}
where 
\begin{eqnarray}
 \fl S[u,\widetilde{u},z,\theta]&=\int_0^td\tau \bigg \{\int_{-\infty}^{\infty}dx\, \widetilde{u}(x,\tau)\partial_{\tau} u(x,\tau)-i\phi\frac{dz}{d\tau} \nonumber \\
 \fl & \quad +z(\tau)\bigg [-\gamma \left (1-\e^{i\phi(\tau)}\right)+\int dx\, \widetilde{u}(x,\tau)\L_{0}u(x,\tau)\bigg ]\nonumber \\
 \fl &\quad +(1-z(\tau))\bigg [-\alpha \left (1-\e^{-i\phi(\tau)}\right)+\int dx\, \widetilde{u}(x,\tau)\L_{1}u(x,\tau)\bigg ]\bigg\}.
\end{eqnarray}

\section{Discussion}

In this paper we derived global density equations for a population of actively switching particles by generalizing the classical formulation of Dean \cite{Dean96}. In the case of a randomly switching environment (extrinsic switching), we showed that the global density $\rho(\x,t)=\sum_{j}\delta(\X_j(t)-\x)$ evolves according to an hSPDE. Averaging with respect to the spatiotemporal white noise process (and using mean-field theory or DDFT in the case of pair-wise interactions), resulted in a hybrid PDE for the 1-particle density $u(\x,t)=\langle \rho(\x,t)\rangle$. We then derived moment equations from the corresponding functional CK equation (\ref{swCK0}), and used this to highlight how statistical correlations are induced by the randomly switching environment, even in the absence of particle-particle interactions. Such correlations are absent when the individual particles independently switch (intrinsic switching). In the latter case we derived a non-hybrid SPDE equation for the indexed set of global densities $\rho_n(\x,t)=\sum_j\delta(\X_j(t)-\x)\E[\delta_{N_j(t)=n}]$. \pcb{However, the inclusion of particle interactions resulted in a moment closure problem for the global densities with respect to the switching process.} Finally, we constructed MSRJD field theoretic formulations of the global density equations in the case of non-interacting particles. 

In this paper, we focused on the derivation and  general mathematical structure of global density equations for actively switching systems. In future specific applications, a number of issues could be explored further.
\medskip

\noindent \pcb{[A] {\em Mean field theory for actively switching particles.} There has been significant progress in the rigorous mathematical analysis of mean-field limits and McKean-Vlasov equations for weakly-interacting particle systems without switching \cite{Jabin17,Oelsch84,Carrillo20}. An extension to a randomly switching environment has also been developed \cite{Nguyen20}. However, as far as we are aware, analogous results for the mean-field limit in the case of intrinsically switching particles has not been considered. As we showed in section 4.2, the inclusion of particle interactions leads to a moment closure problem at the level of the generalized Dean-Kawasaki equation for the global densities, see for example equation (\ref{drhoc2}). The effects of environmental switching on  the validity of the adiabatic approximation for DDFT also needs to be investigated. In the fast switching limit one could average with respect to switching and then apply DDFT, and vice versa in the slow switching limit. The difficulty arises at switching rates comparable to the rate of relaxation to thermodynamic equilibrium.}
\medskip

\noindent [B] {\em Coupling between the white noise and switching processes.} In the case of particle switching, we assumed that the diffusivity $D$ of the $j$th particle is independent of its discrete internal state $N_j(t)$. However, advances in single-particle tracking (SPT) and statistical methods suggest that particles within the plasma membrane, for example, can switch between different discrete conformational states with different diffusivities \cite{Das09,Persson13,Slator15}. Such switching could be due to interactions between proteins and the actin cytoskeleton or due to protein-lipid interactions. Interestingly, the switching rates between the different conformational states could also depend on the spatial location of a particle. 
For example, an experimental and computational study of {\em C.~elegans} zygotes showed that protein concentration formation during cell polarization relies on a space-dependent switching mechanism \cite{Wu18}. This was independently predicted in a general theoretical study of protein gradient formation in switching systems \cite{Bressloff17b,Bressloff19}. For space-dependent switching we have to modify the definition (\ref{dglob}) according to 
\[ \rho_n(\x,t)=\sum_{j=1}^{\calN}\E[\delta_{N_j(t),n}\delta(\X_j(t)-\x)],\]
and equation (\ref{drhocav}) becomes
\begin{eqnarray}
\fl \frac{\partial \rho_n(\x,t)}{\partial t} 
&=\sqrt{2D_n}{\bm \nabla} \cdot \bigg [ \sqrt{\rho_n(\x,t)} {\bm \eta}_n(\x,t)\bigg ]+D_n{\bm \nabla}^2  \rho_n(\x,t) 
-{\bm \nabla} \cdot  [\rho_n(\x,t){\bf A}_n(\x)] \nonumber \\
\fl &\quad +\sum_{m=0}^{K-1}Q_{nm}(\x)\rho_m(\x,t).
\end{eqnarray}

\begin{figure}[t!]
\centering
\includegraphics[width=7cm]{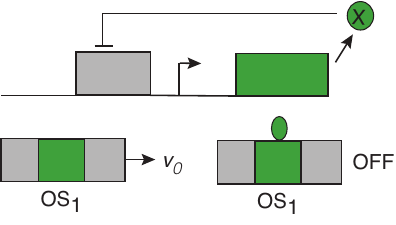}
\caption{An autoregulatory gene network with single operator site $OS_1$. A gene is repressed (or activated) by its own protein product $X$.}
\label{fig4}
\end{figure}

\noindent [C] {\em $X$-dependent switching in regulatory gene networks.} Another important example of an actively switching system with a state-dependent matrix generator ${\bf Q}$ is a gene network with regulatory feedback \cite{Kepler01}. The simplest type of a feedback circuit involves a gene that is regulated by its own protein product (auroregulation), as shown in Fig. \ref{fig4}. Suppose that the promoter has a single operator site $OS_1$ for binding protein X. The gene is assumed to be 
OFF when $X$ is bound to the promoter and 
ON otherwise. If $O_0$ and $O_1$ denote the unbound and bound 
promoter states, then the corresponding reaction scheme is
\[O_0  \Markov{\alpha}{\beta x}  O_1,\]
where $x$ is the concentration of $X$. The concentration evolves according to the piecewise deterministic equation
\begin{equation}
\label{geneCK}
\frac{dx}{dt}=A_n(x),\quad  \mbox{ for } N(t)=n,
\end{equation}
where $A_0(x)=\gamma_0-\kappa_0x$, $A_1(x)=-\kappa_0 x$
and discrete state transitions are generated by the matrix
\begin{equation}
{\bf Q}(x)= \left (\begin{array}{cc} -\gamma x  &\quad \alpha 
\\ \gamma x& \quad -\alpha  \end{array}
\right ).
\label{autoQ}
\end{equation}

\noindent [D] {\em Weak noise limit.} Another important application of field theory is to the derivation of least action principles in the weak noise limit. For the given population model, there are two distinct sources of noise at the single particle level: (i) the stochastic switching between different internal states; (ii) external white noise with diffusivity $D$. The weak noise limit involves taking $D\rightarrow 0$ and \pcb{ $W_{nm} \rightarrow \infty$}. A typical feature of a stochastic path integral is that the sum-over-paths has support over the set of paths that are continuous but non-differentiable with respect to $\tau$. In particular, any time derivative in the action functional is a formal symbol for the appropriate difference term in the time-discretized path integral.  Nevertheless, in the weak noise limit, the path integral is dominated by paths that are arbitrarily close to the classical least action paths, which are differentiable.

\end{document}